\documentclass[12pt, draftclsnofoot, onecolumn]{IEEEtran}

\renewcommand{\IEEEbibitemsep}{0pt plus 0pt}
\makeatletter
\IEEEtriggercmd{\reset@font\normalfont\footnotesize}
\makeatother
\IEEEtriggeratref{1}

\usepackage{etex}

\usepackage{algpseudocode}
\usepackage{algorithm}
\usepackage{amsmath}
\usepackage{gensymb}
\usepackage{amsthm}
\usepackage{mathtools}

\newcommand{\vc}[1]{{\mathbf{#1}}}
\newcommand{\comment}[1]{}

\renewcommand{\Comment}[2][.35\linewidth]{%
  \leavevmode\hfill\makebox[#1][l]{//~#2}}
\algnewcommand\algorithmicto{\textbf{to}}
\algnewcommand\RETURN{\State \textbf{return} }

\usepackage{bm}
\usepackage{amssymb}

\usepackage{varwidth}
\DeclareMathOperator*{\argmax}{arg\,max}
\usepackage[noadjust]{cite}

\usepackage{enumerate}
\usepackage{amssymb}
\usepackage{multirow}
\usepackage{dsfont}
\usepackage{citesort}
\usepackage{array}
\newenvironment{definition}[2][Definition]{\begin{trivlist}
\item[\hskip \labelsep {\bfseries #1}\hskip \labelsep {\bfseries #2}]}{\end{trivlist}}
\newtheorem{lemma}{Lemma}
\newtheorem{theorem}{Theorem}

\usepackage{listings}
\usepackage[utf8x]{inputenc} 
\usepackage{graphicx}
\usepackage{float}
\usepackage{epstopdf}
\usepackage{bbm}
\usepackage{pbox}
\usepackage{color}
\usepackage[section,subsection,subsubsection]{extraplaceins}
\usepackage{multirow}
\usepackage{subfigure}
\usepackage{booktabs}

\newcommand{\RNum}[1]{\uppercase\expandafter{\romannumeral #1\relax}}
\newcommand{\suchthat}{\;\ifnum\currentgrouptype=16 \middle\fi|\;}

\usepackage{amsmath}
\usepackage{tikz}
\usetikzlibrary{automata,arrows,positioning,calc}

\usetikzlibrary{arrows}
\tikzstyle{int}=[draw, fill=blue!20, minimum size=2em]
\tikzstyle{init} = [pin edge={to-,thin,black}]

\usetikzlibrary{positioning,chains,fit,shapes,calc, arrows, shadows,trees}
\usepackage{calc}
\usepackage{multicol}
\usepackage{lipsum}
\usetikzlibrary{decorations.markings, arrows, automata}
\tikzstyle{vertex}=[circle, draw, inner sep=0pt, minimum size=5pt]
\tikzset{symbol/.style={rectangle, draw, very thick,
minimum size=10mm, rounded corners=1mm}}
\tikzset{symbol2/.style={rectangle , draw,  thick,
minimum size=35mm, rounded corners=1mm}}

\tikzstyle{block} = [draw, fill=white, rectangle, 
    minimum height=3em, minimum width=6em]
\tikzstyle{sum} = [draw, fill=gray, circle, node distance=1cm]
\tikzstyle{input} = [coordinate]
\tikzstyle{output} = [coordinate]
\tikzstyle{pinstyle} = [pin edge={to-,thick,black}]

\begin{document}
\title{\LARGE{Dual Sub-6 GHz -- Millimeter Wave Beamforming and Communications to Achieve  Low Latency and High Energy Efficiency in 5G  Systems}}
\author{\IEEEauthorblockN{Morteza Hashemi\IEEEauthorrefmark{1}, C. Emre Koksal\IEEEauthorrefmark{1}, and Ness B. Shroff \IEEEauthorrefmark{1}\IEEEauthorrefmark{2}}\\\IEEEauthorblockA{\IEEEauthorrefmark{1}  Department of Electrical and Computer Engineering, The Ohio State University} \\
\IEEEauthorblockA{\IEEEauthorrefmark{2}Department of Computer Science and Engineering, The Ohio State University}}

\maketitle

\begin{abstract}
We propose an architecture that integrates RF (i.e., sub-6 GHz) and millimeter wave (mmWave)  technologies for $5$G cellular systems.
Communications in the mmWave band faces significant challenges due to variable channels, intermittent connectivity, and high energy usage. Moreover, speeds for electronic processing of data is of the same order as typical rates for mmWave interfaces which makes the use of complex algorithms for tracking channel variations and adjusting resources accordingly impractical. 
Our proposed architecture integrates the RF and mmWave interfaces for \emph{beamforming} and \emph{data transfer}, and  exploits the spatio-temporal correlations between the interfaces. Based on extensive experimentation in indoor and outdoor settings, we demonstrate that an integrated RF/mmWave signaling and channel estimation scheme can remedy the problem of high energy usage and delay associated with mmWave beamforming. In addition, cooperation between two interfaces at the higher layers effectively addresses the high delays caused by highly intermittent mmWave connectivity. We design a scheduler that fully exploits the mmWave bandwidth, while the RF link acts as a fallback mechanism to prevent high delay. To this end, we formulate an optimal scheduling problem over the RF and mmWave interfaces where the goal is to maximize the delay-constrained throughput of the mmWave interface. We prove using subadditivity analysis that the optimal scheduling policy is based on a single threshold that can be easily adopted despite high link variations.  
\end{abstract}

\begin{IEEEkeywords}
Millimeter wave communication, 5G mobile systems, Out-of-band beamforming and communication
\end{IEEEkeywords}

\section{Introduction}
The annual data traffic generated by mobile devices is expected to surpass $130$ exabits by 2020 \cite{khan2011mmwave}. This deluge of traffic will significantly exacerbate the spectrum 
crunch that cellular providers are already experiencing. To address this issue, it is envisioned that in $5$G cellular systems certain portions of the mmWave band will be used, spanning the spectrum between $30$ GHz to $300$ GHz with the corresponding wavelengths between $1$-$10$ mm \cite{rappaport2013millimeter}. This will substantially increase the spectrum available to cellular providers, which is currently between $700$ MHz and $2.6$ GHz with only $780$ MHz of bandwidth allocation for all current cellular technologies. However, before mmWave communications can become a reality, there are significant challenges that need to be overcome.

Firstly, compared with RF (i.e., sub-6 GHz), the propagation loss in mmWave is much higher due to atmospheric absorption and low penetration. Although large antenna arrays have the potential to make up for the mmWave losses, they cause several other issues such as high energy consumption by components (e.g., analog-to-digital converters (ADC)).
For instance, at a sampling rate of $1.6$ Gsamples/sec, an $8-$bit quantizer consumes $\approx 250$mW of power.
During active transmissions, this would constitute up to $50\%$ of the overall power consumed for a typical
smart phone. Moreover, in order to fully utilize the  directional antenna arrays, continuous beamforming and signal training at the receiver is needed \cite{roh2014millimeter}. Digital beamforming is highly efficient in delay, but there is a need for a separate ADC for each antenna, which may not be feasible for even a small to mid-sized antenna array due to high energy consumption. In contrast, analog beamforming requires only one ADC, but it can focus on one direction at a time, making the angular search costly in delay.  There are also proposals on hybrid digital/analog beamforming, which strikes a balance between analog and digital beamforming, using a few ADCs rather than one per antenna \cite{mo2016hybrid}.

Secondly, mmWave channels can be highly variable with intermittent connectivity since most objects lead to blocking and reflections as opposed to scattering and diffraction in typical RF frequencies. When the users and/or surrounding objects are mobile, different propagation paths become highly variable with intermittent on-off periods, which can potentially result in long outages and poor mmWave delay performance.  On the other hand, ultimately, the very-high bandwidth available in the mmWave band should translate into performance guarantees, required by next generation real-time applications that are expected to dominate the traffic in the next generation networks. 



In order to address the high energy consumption by components and beamforming overhead in mmWave, we consider an RF-assisted mmWave beamforming scheme. Due to high cost and
energy consumption by ADCs in fully-digital beamforming as well as the delay in fully-analog
beamforming, we investigate the feasibility of conducting a coarse angle of arrival (AoA) estimation on the RF channel and then utilizing the fully-analog
beamforming for fine tuning and transmissions.  To this end, we first experimentally verify the correlation between the RF and mmWave AoA,
especially in the presence of line-of-sight (LOS). Our measurements taken jointly at different bands and for both \emph{indoor and outdoor settings} show that under LOS conditions and in $94\%$ of all measurements, the identified AoA of signal in the RF band is within $\pm 10\degree$ accuracy for the AoA of the mmWave signal.  Based on the estimated RF AoA, the angular range over which we scan for the mmWave transmitter reduces to no more than $20\degree$ on average, from $180\degree$ in stand-alone mmWave systems. 
Note that the authors in \cite{nitsche2015steering} also proposed a beamforming method based on out-of-band measurements for $60$~GHz WiFi and under \emph{static indoor} conditions.

To mitigate the issue of highly intermittent mmWave connectivity, we consider an integrated RF/mmWave transceiver model, in which, in addition to beamforming, the RF interface is used for data transfer.  The link speed of the mmWave
interface (multi-Gbps) is comparable to the speed at which a typical processor in a smart device operates. This is different from classical wireless interfaces in which data
rates are much smaller than the clock speeds of the processors.
Thus, the mmWave interface cannot be assumed to operate at smaller time-scales and the
algorithms run at the processor may not be able to respond to variations in real time and execute control decisions. \emph{This necessitates the
use of proactive queue-control solutions along with a reasonably large buffer at the mmWave interface.} For instance, if the queue size at the mmWave interface gets small, the risk of wasting the abundant capacity from mmWave increases. Conversely, if we keep the queue at the mmWave interface large, if the channel goes down, we incur a high delay.  

To understand the tradeoff between full exploitation of the mmWave capacity and delay for the mmWave channel access,  \emph{we model the RF/mmWave transceiver as a communication network, and investigate an optimal scheduling policy using network optimization tools. } Specifically, in the equivalent network model, the RF and mmWave interfaces are represented by individual network nodes with dedicated queues. Hence, the optimal transmission policy across the RF and mmWave interfaces is transformed into an optimal scheduling policy across the RF and mmWave nodes. 
We formulate an optimal scheduling problem where the objective is to achieve maximum mmWave channel utilization with bounded delay performance. In order to determine ``when'' a data packet should be added to the RF or mmWave queues, we prove that the optimal policy is of the \emph{threshold-type} such that the scheduler routes the arrival traffic to the mmWave queue if and only if its queue length is smaller than a threshold. We show that the threshold-based scheduling policy efficiently captures the dynamics of the mmWave channel, and indeed maximizes the channel utilization.  In summary, our main contributions are as follows:
\begin{itemize}
\item We have conducted a wide variety of experiments to evaluate the correlation between the measured channel gains for the $30$ GHz mmWave and $3$ GHz RF interfaces under various indoor and outdoor situations involving existence of LOS between the transmitter and receiver. 

\item We propose an integrated RF/mmWave system that exploits the cross-interface correlations for beamforming as well as data transfer. Our ADC follows the beamformer at the receiver and eliminates the need for a separate ADC for each element in the mmWave antenna-array.

\item We propose a framework to model the integrated RF/mmWave transceiver as a network and jointly manage the transmission across the RF and mmWave interfaces. Our queue management formulation explicitly takes into account the mmWave channel dynamics, and our approach enables full utilization of the available mmWave channel capacity, despite the highly variable nature of the channel. We prove using subadditivity analysis that the optimal scheduling policy is a simple threshold-based one, which can be easily adopted despite the high link variations.
\end{itemize}
We should emphasize that the RF/mmWave correlation was studied in \cite{aliestimating}, and applied only for beamforming  in \cite{nitsche2015steering}. However, a coherent design that fully integrates the RF and mmWave interfaces and optimally design the RF/mmWave transceivers is missing. Hence, we aim to develop an integrated architecture for which the RF interface is utilized for both beamforming and data transfer. A preliminary version of our results was presented in \cite{wiopt-2017}. 

\subsection{Notations} We use the following notation throughout the paper. Bold uppercase and lowercase letters are used for matrices and vectors, respectively, while non-bold letters are used for scalers.  In addition, $(.)^\dagger$ denotes the conjugate transpose, $\text{tr}(.)$ denotes the matrix trace operator, and $\mathds{E}[.]$ denotes the expectation operator. The RF and mmWave variables are denoted by $(.)_\text{RF}$ and $(.)_\text{mm}$, respectively.

\section{Related Work}  
We classify existing and related work across the following thrusts: 

\subsection{Experimental Studies} 
Wireless channel fading is primarily studied under two disparate categories based on the impact and the
time-scale of the associated variations: large-scale (due to shadowing, path loss, etc.) and small-scale (due
to mobility combined with multipath). 
There exist numerous  measurement and experimentation efforts in order to understand mmWave propagation and  the effect of slow scale and large scale fading in the mmWave band (see, for example, \cite{rappaport2013millimeter,collonge2004influence}). The main objective has been to extend the existing far-field ray-tracing models to accurately represent various phenomena observed in mmWave. For example, in~\cite{rangan2014millimeter,rappaport2014millimeter}, a model based on isolated clusters is argued to be more appropriate to capture the observed reflections in mmWave, as opposed to the uniform distribution across the delay taps. Extensive evaluations of mmWave propagation taken from hundreds of different locations and settings also exist, by the same group~\cite{rappaport2013millimeter,rappaport2014millimeter,rangan2014millimeter} as well as others~\cite{nurmela2015metis}. Our goal is to neither replicate nor expand these observations. Instead, we are interested in the channel/propagation environment correlation across different interfaces under various conditions, including indoor and outdoor situations, with mobility, and existence of LOS.

\subsection{MmWave Beamforming and Communications}
There has been extensive amount of work on digital and analog beamforming methods  (e.g., \cite{roh2014millimeter,adhikary2014joint}). There are also proposals on hybrid beamforming methods \cite{mo2016hybrid} in which the term ``hybrid'' refers to the
mixture of analog/digital (different from our hybrid RF/mmWave system). The whole operation there is in
sole mmWave domain.  Recently, there have been proposals on leveraging out-of-band information in order to enhance the mmWave performance. The authors in \cite{aliestimating} propose a transform method to translate the spatial correlation matrix at the RF band into the correlation matrix of the mmWave channel. The authors in \cite{nitsche2015steering} consider the $60$ GHz indoor WiFi network, and investigate the correlation between the estimated AoA at the RF band with the mmWave AoA in order to reduce the beam-steering
overhead. The authors in \cite{ali2017millimeter} propose a compressed beam selection method which is based on out-of-band spatial information obtained at the sub-6 GHz band. In the context of hybrid communications and data transfer, the authors in \cite{kim2014analytical} studied a dual interface system to offload cellular data over WiFi network.  In another line of research, there are proposals to integrate 3G, WiFi and WiMAX \cite{jindal2005grouping}. There have also been some recent studies \cite{nitsche2014ieee,shokri2015design,nitsche2015boon,yildirim2009cross,mezzavilla20155g} of potential approaches for higher layer design in mmWave networks. These approaches mainly focus on multiple access schemes, given the unique propagation characteristics of the medium. 

Our work is distinguished from the previous work as (i) we experimentally investigate the RF/mmWave correlation under practical scenarios, and demonstrate that how mobility affects the channel conditions and cross-interface correlation, and (ii) we propose a holistic RF/mmWave architecture wherein the RF interface is exploited for beamforming as well as data transfer in order to reduce the energy consumption and prevent high delay caused by mmWave outages. To the best of our knowledge, there is no previous work that considers an integrated RF/mmWave architecture for joint beamforming and data transfer with optimal scheduling policy. 

\section{Integrated RF/mmWave Architecture}
\label{architecture}
\subsection{System Model}

\begin{figure*}[t!]
 \vspace{-.5cm}
\begin{center}
\includegraphics[height=1.3in]{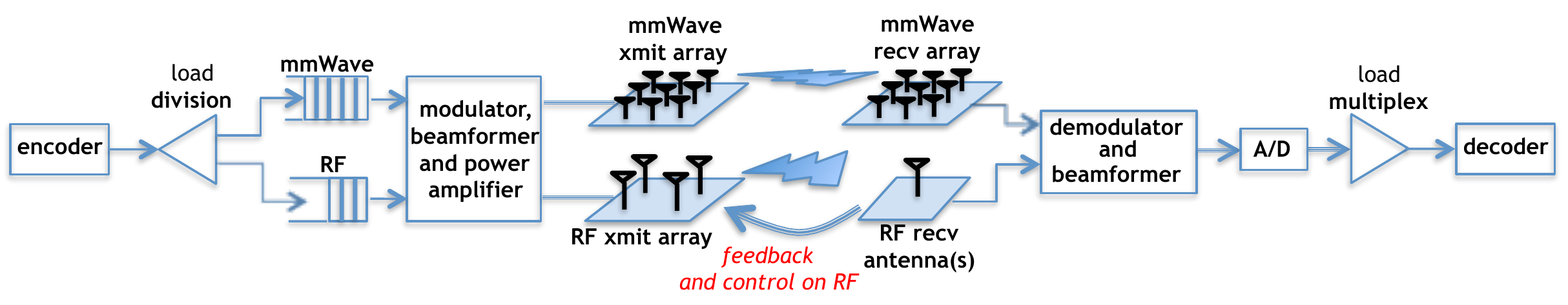}
\vspace{-0.1in}
\caption{\small{Our hybrid communications system. The speed of mmWave interface necessitates the use of a separate queue at the input of that interface. RF antenna arrays are needed at the access points and not necessary at the mobiles.
}}
\vspace{-0.25in}
\label{fig:system}
\end{center}
\end{figure*}

Figure \ref{fig:system} illustrates the basic components of our proposed architecture that exploits the correlation between the mmWave and RF channels as it pertains to large-scale effects and AoA in the presence of LOS path \cite{nitsche2015steering,aliestimating}. The proposed architecture addresses the energy issue by:
\begin{itemize}
\item Exploiting the cross-interface correlation to achieve the beamforming fully in the analog domain. Thus,
the ADC follows the beamformer at the receiver, and eliminates the need for a separate one, for all elements in the mmWave antenna-array. 

\item Moving all mmWave control signaling and channel state information (CSI) feedback to the RF interface, and thus avoiding the two-way beamforming and reverse channel transmission costs in mmWave.
\end{itemize}

In addition to high energy consumption by components,  the mmWave channel is highly sensitive and outages can be long that can lead to unacceptably high delays for delay-sensitive applications. However, a conservative use of the mmWave link is not desirable either, since the upside of the mmWave channel can be enormous,
especially in the presence of LOS that occurs intermittently. More importantly, the high data rate of the mmWave link necessitates the use of a reasonably large buffer at the mmWave interface along with proactive queue-control solutions. In this context, we derive an optimal scheduling policy to select which interface(s) to use and control the queue sizes of interfaces in order to achieve maximum mmWave throughput with guaranteed constrained delay. We investigate the optimal interface scheduling in Section \ref{rf-assisted-communication}.

\begin{figure}[t]
\centering
\includegraphics[scale=.4, trim = 0cm 1.5cm 0cm 2.8cm, clip]{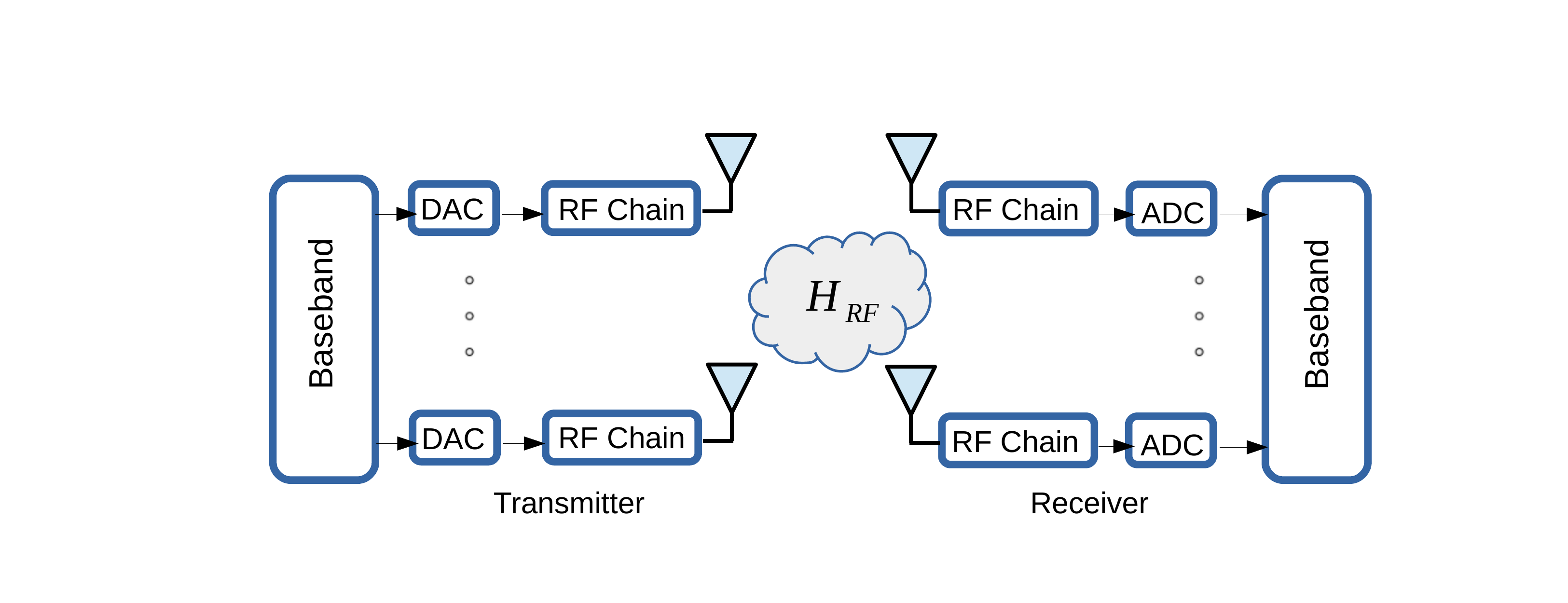}
\caption{RF model based on digital beamforming.}
\label{fig:RF-model}
\end{figure}

\subsection{RF System and Channel Model}
For the RF system, we use digital beamforming  as shown in Fig. \ref{fig:mmWave-model}. As a result, the received signal at the receiver can be written as: 
\begin{equation}
\label{eq:rec_signals}
\mathbf{y}_{\text{RF}}=\mathbf{H}_{\text{RF}} \cdot
\mathbf{x}_{\text{RF}}+\mathbf{n}_{\text{RF}},
\end{equation}
where $\mathbf{H}_{\text{RF}}$ is the RF-channel matrix and $\mathbf{x}_{\text{RF}}$ is the transmitted signal vector in RF. Entries of circularly symmetric white Gaussian noise are denoted by $\mathbf{n}_{\text{R}}$. The RF receiver uses the steering vector $\vc{w}_{\theta_\text{RF}}$ to align the received signals where the optimal steering direction ${\theta^*_\text{RF}}$ can be obtained based on maximizing the SNR, i.e.,:
\begin{equation}
\theta^*_\text{RF} = \argmax_{{\theta_\text{RF}}} \ \frac{\vc{w}_{\theta_\text{RF}}^\dagger \mathbf{H}_{\text{RF}} \mathbf{K}_{\mathbf{xx}} \mathbf{H}_{\text{RF}}^\dagger\vc{w}_{\theta_\text{RF}}}{N_0},
\end{equation}
in which $\mathbf{K}_{\mathbf{xx}}$ is the covariance matrix, and $N_0$ is the noise power.

\subsection{MmWave System and Channel Model} 
\begin{figure}[t]
\centering
\includegraphics[scale=.4, trim = 0cm 1.2cm 0cm 1.5cm, clip]{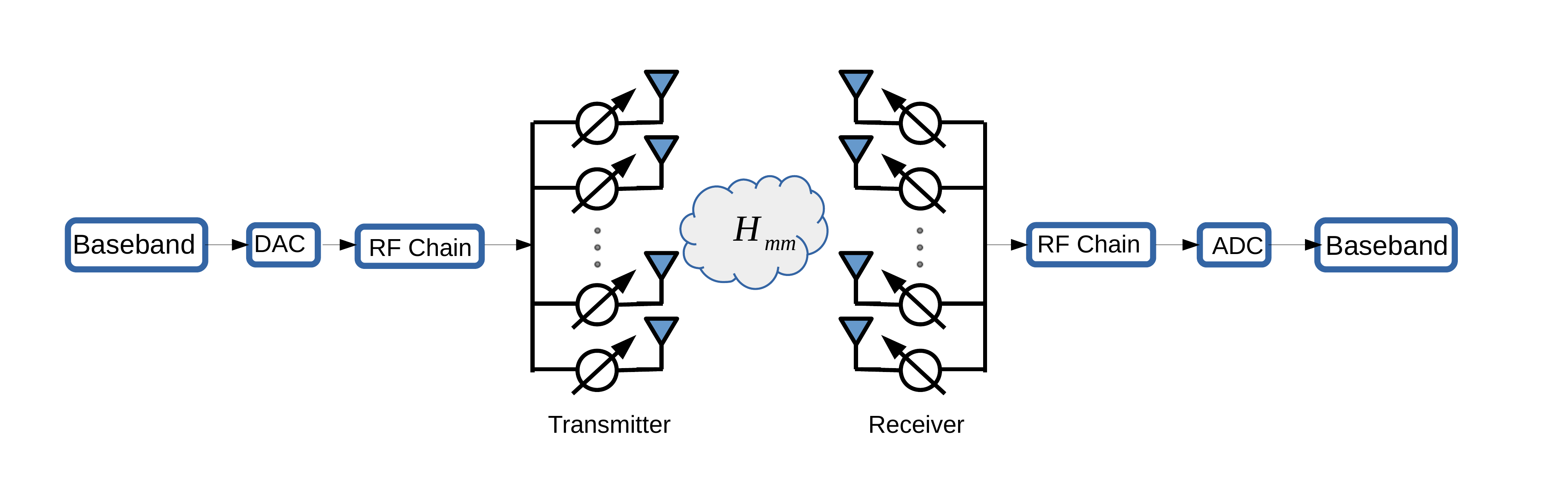}
\caption{MmWave system model based on analog beamforming.}
\label{fig:mmWave-model}
\end{figure}

 The mmWave system model is shown in Fig. \ref{fig:mmWave-model}. Unlike RF, we use  analog combining for mmWave via a single ADC to avoid high energy consumption.  Consequently, the signal at the input of the decoder is a scalar, identical to a weighted combination of signal $x_{\text{mm}}$ across all antennas. Thus, the received signal at the mmWave receiver can be written as: 
\begin{equation}
\label{eq:mm_rec_signals}
y_{\text{mm}}=\vc{w}_r^\dagger \mathbf{H}_{\text{mm}} \vc{w}_t \cdot x_{\text{mm}}+n_{\text{mm}}, 
\end{equation}
where $\vc{w}_r$ and $\vc{w}_t$ are the analog-receive and digital-transmit beamforming vectors, and $n_{\text{mm}}$ denotes the white Gaussian noise component.  \emph{Note that, our formulation can readily be extended to the case with digital combining at mmWave, in case ADC conversion is made at the output of each antenna}.

In the mmWave domain, the channel matrix $\mathbf{H}_\text{mm}$ has a singular value decomposition
$
\mathbf{H}_\text{mm} = \mathbf{U} \mathbf{\Lambda} \mathbf{V^\dagger}, 
$ 
where $\mathbf{U} \in \mathcal{C}^{n_r \times n_r}$ and $\mathbf{V} \in \mathcal{C}^{n_t \times n_t}$ are rotation unitary matrices and $\mathbf{\Lambda} \in \mathcal{R}^{n_r \times n_t} $ is a diagonal matrix whose diagonal elements are nonnegative real numbers $\rho_1 \geq \rho_2 \geq ... \geq \rho_{n_\text{min}} $, where $n_\text{min} = \min (n_r, n_t)$, and $n_t$ and $n_r$ denote the number of mmWave transmit and receive antennas, respectively. The mmWave-channel matrix $\mathbf{H}_\text{mm}$ is low rank \cite{andrews2016modeling}, and since the rank of $\mathbf{H}_\text{mm}$ is equal to the number of non-zero singular values, we restrict our attention to only the largest eigenvalue $\rho_1$ and assume that $\rho_1 \gg \rho_i$, and that $\rho_i\approx0$ for $\ i \neq 1$. In fact, our experimental results show that under the LOS conditions, there is about $10 - 15$ dB gain improvement due to the strongest eigenmode, and thus we assume that the state of mmWave link can be characterized based on the value of $\rho_1$. Given the assumed channel models, next we experimentally investigate the correlation between the RF and mmWave channels under various conditions.

\subsection{Integrated RF/mmWave Experimentation} 
\begin{figure*}[t!]
\vspace{-0.2in}
  \begin{center}
    \subfigure[Basic setup for indoor experiments] {
      \label{fig:in_setup}
      \includegraphics[height=1.3in]{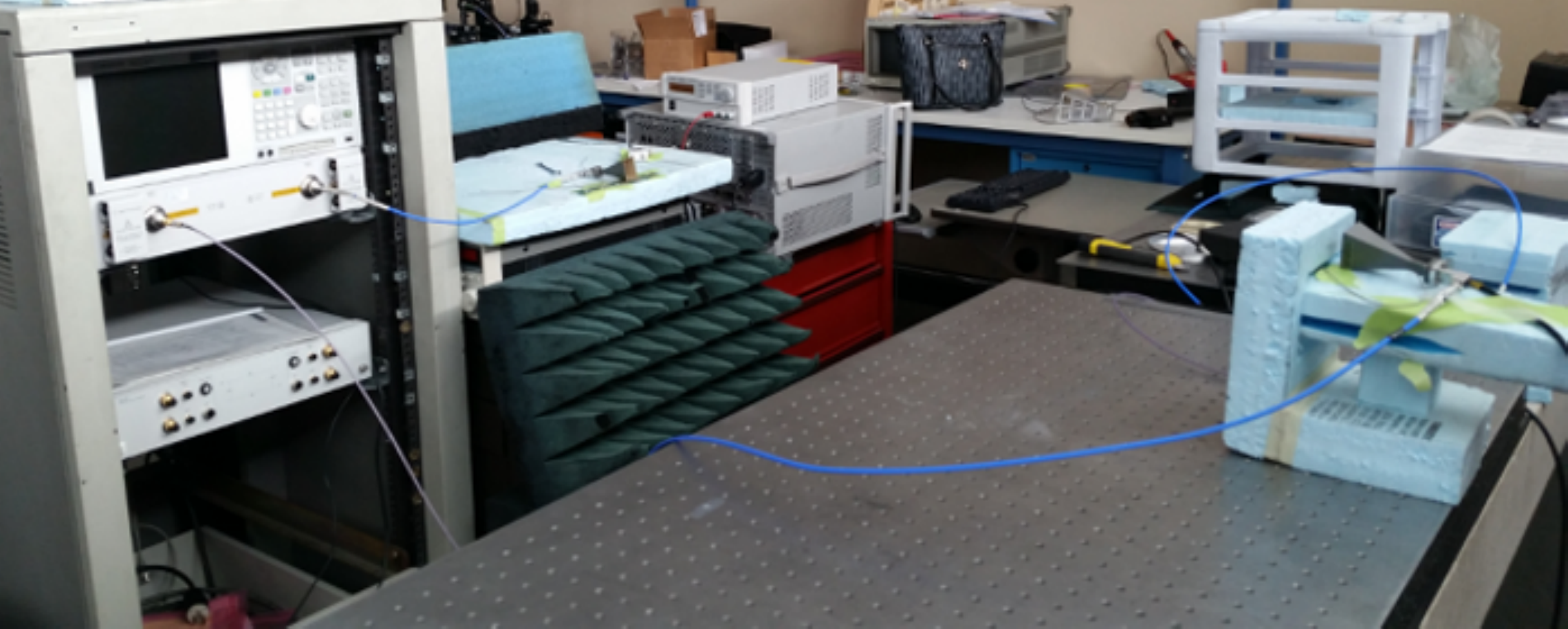}  \hspace{0.4in}
    } 
    \subfigure[Basic setup for outdoor experiments] {
      \label{fig:out_setup}
      \includegraphics[height=1.3in]{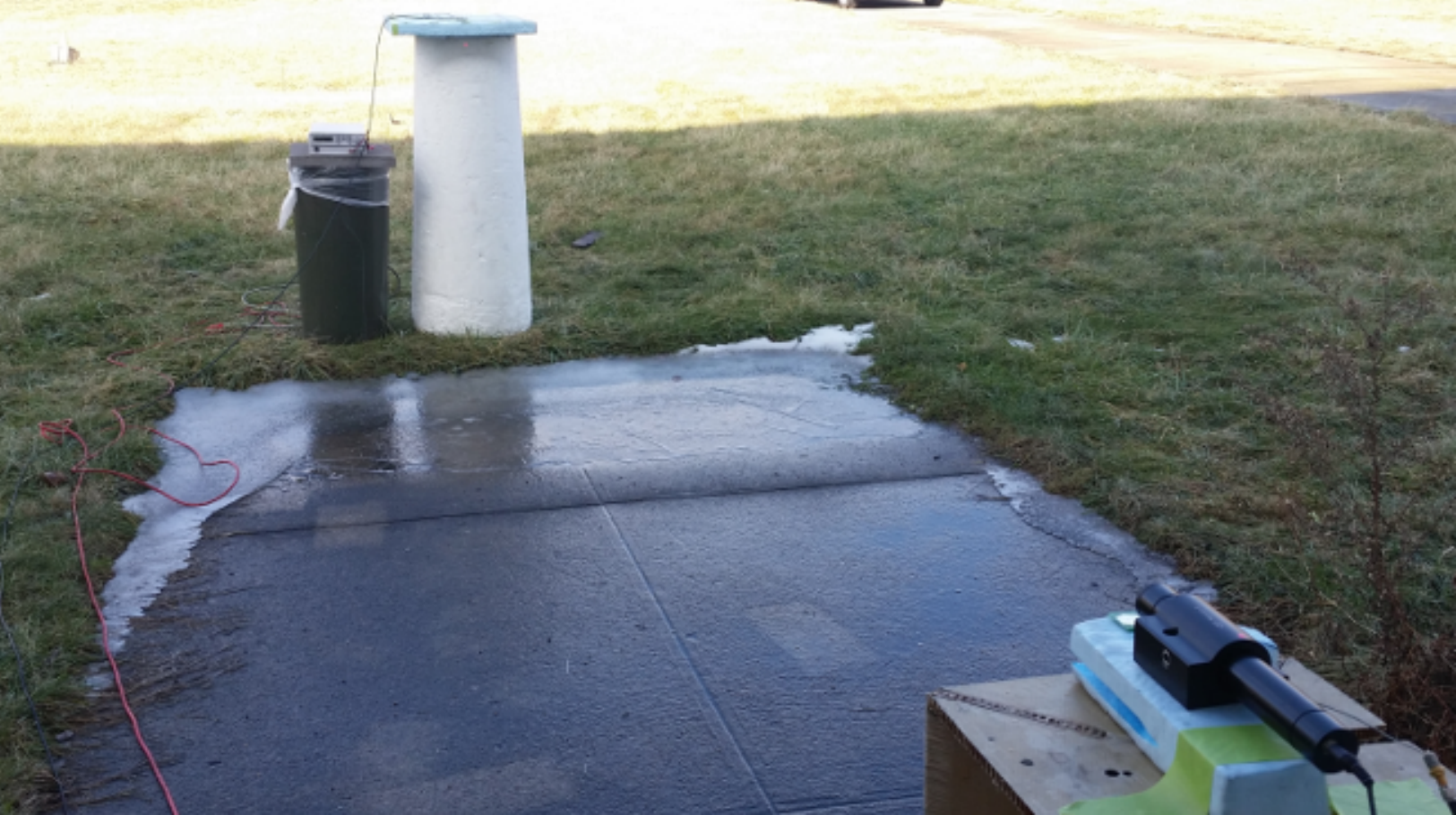}  
    }
  \end{center}
  \vspace{-0.25in}
  \caption{{We use a joint mmWave and RF measurement setup to observe various properties of the propagation environment jointly across the two bands. The setup involves a network analyzer, horn antennas for mmWave, and omnidirectional 4-antenna array for RF measurements.}}
  \vspace{-0.2in}
  \label{fig:system_and_scheme}
\end{figure*}

\textbf{Experimental setup:} We simultaneously observe the RF and mmWave channels via a dual transmitter-receiver pair in the same location. Our experimental setup is shown in Fig. \ref{fig:system_and_scheme}. In the RF platform, we use an omni-directional antenna operating at
$3$~GHz as a transmitter and $4$  omni-directional antennas as a receiver in order to observe the AoA for the incoming RF signal.  
 We use the MUSIC  algorithm\footnote{For the sake of clarity, we use MUSIC algorithm, but other estimators can be used as well.} \cite{schmidt1986multiple} to evaluate the components of the signals across various angles. For mmWave, we use $30$ GHz directional antennas to be able to align the beams. We measure the channel across the $180\degree$ space with $10\degree$ step size. Based on a large set of measurements, we conclude that the propagation situations can be classified into three types as it pertains to summarizing the connection between the large-scale effects in RF and mmWave: line-of-sight (LOS), blocker, and non line-of-sight (NLOS). LOS implies that there is a strong line of sight path between the transmitter and the receiver; blocker indicates that, the LOS path for the mmWave interface is being blocked by a non-stationary obstacle; and NLOS indicates the presence of a stationary obstacle, unlikely to change in time. 

   \begin{figure*}[t!]
\vspace{-0.5in}
  \begin{center}
    \subfigure[RF and mmWave activity vs. angle in {\em indoor} setting] {
      \label{fig:indoor_measurements}
      \includegraphics[height=3.2in]{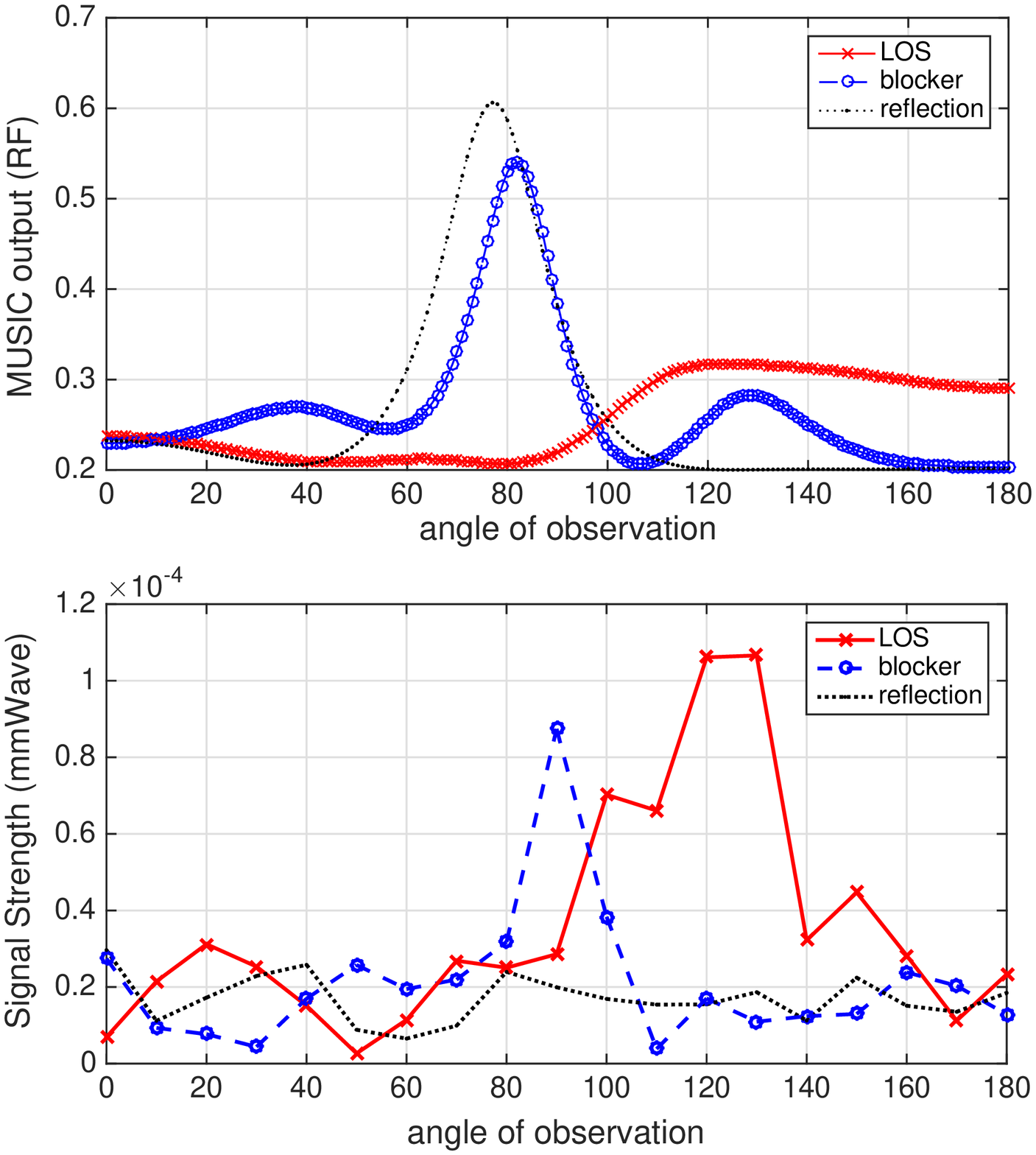}  \hspace{0.1in}
    } 
    \subfigure[RF and mmWave activity vs. angle in {\em outdoor} setting] {
      \label{fig:outdoor_measurements}
      \includegraphics[height=3.2in]{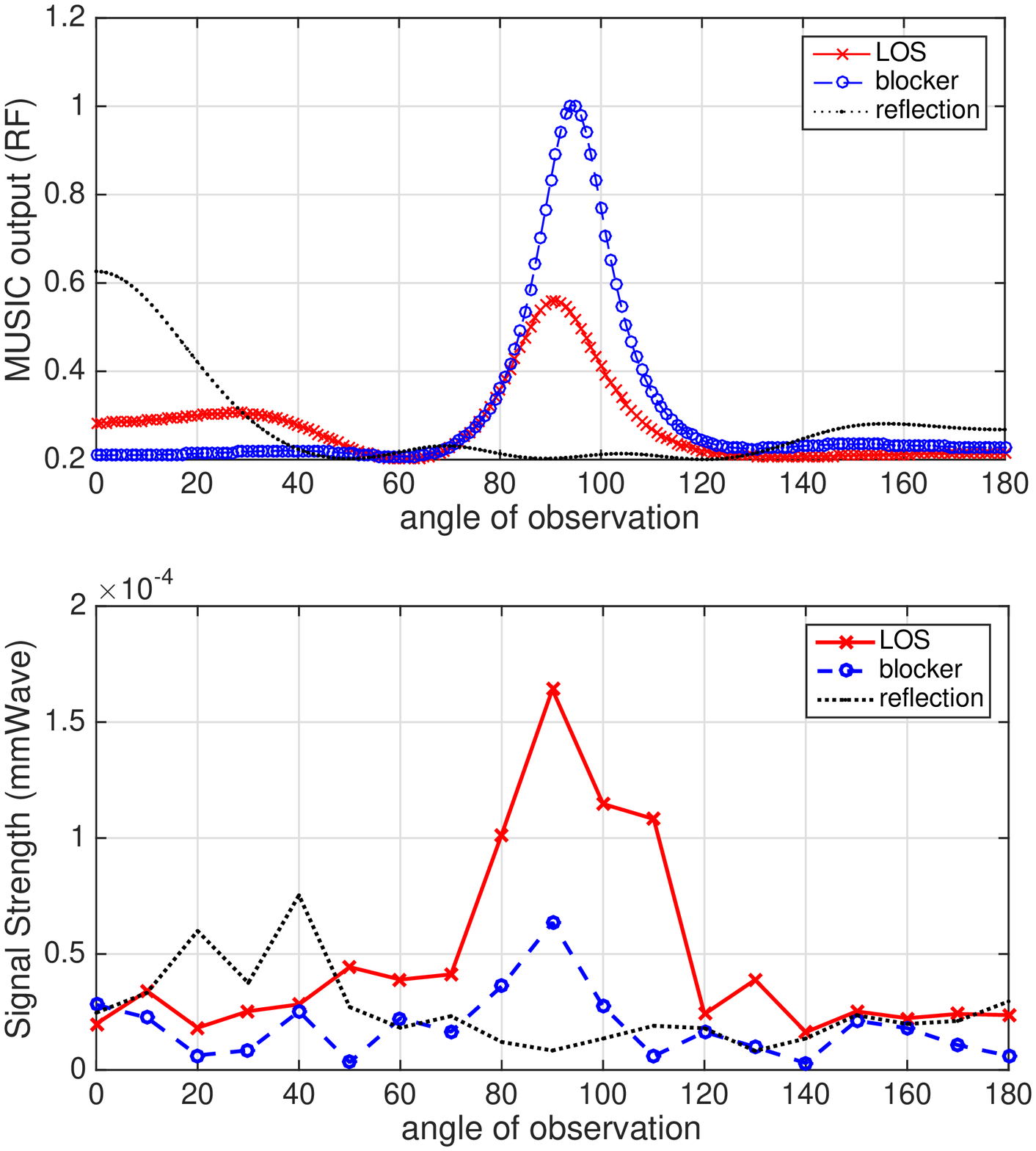}  
    }
  \end{center}
  \vspace{-0.15in}
  \caption{\small{Indoor (\ref{fig:indoor_measurements}) and outdoor (\ref{fig:outdoor_measurements}) associated with RF (top plots) and mmWave (bottom plots). In each case, we have tested three situations: LOS, blocker, and NLOS with reflector.  The direction of strong signal is highly correlated between RF and mmWave if a LOS is present. The correlation is lost in part, if there is a blocker present and lost completely in the case of NLOS with reflections.}}
  \vspace{-0.2in}
  \label{fig:indoor_outdoor_measurements}
\end{figure*}

\begin{figure}[t!]
  \begin{center}
    \subfigure[RF activity vs. angle in indoor setting] {
      \label{RF_HB}
      \includegraphics[scale=.23]{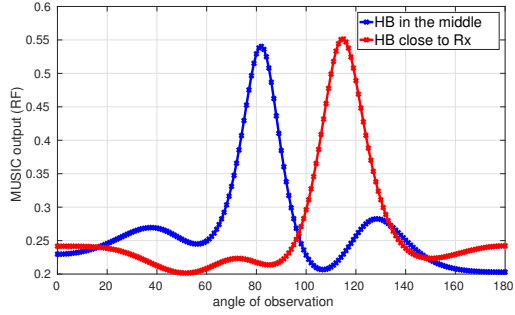} 
    } 
    \subfigure[mmWave activity vs. angle in indoor setting] 
    {
      \label{mmWave_HB}
      \includegraphics[scale=.23]{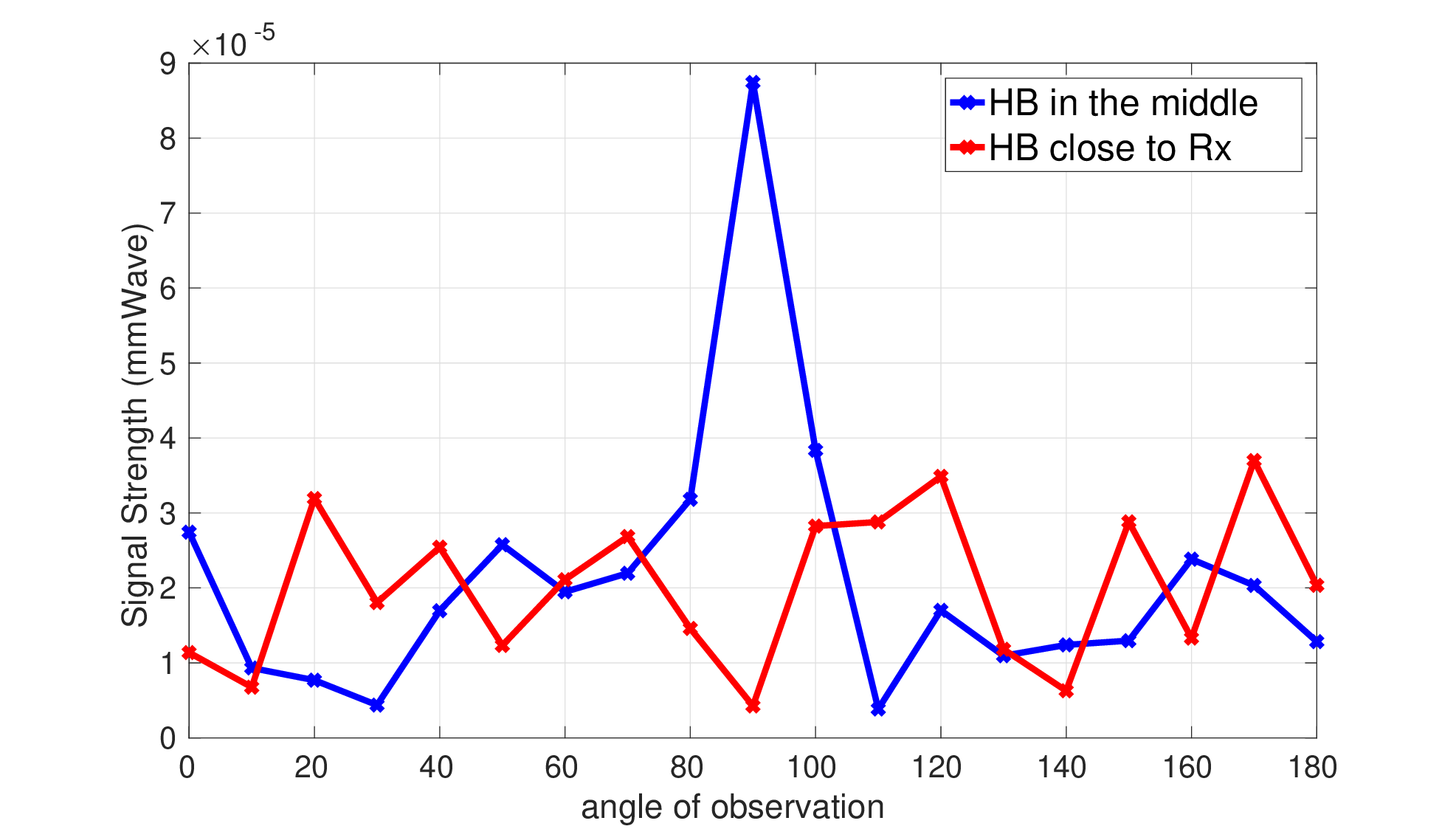}  
    }
  \end{center}
  \vspace{-0.15in}
  \caption{\small{Channel spatial behavior for human block (HB) in indoor environment for RF (top) and mmWave (bottom). }}
  \vspace{-0.2in}
  \label{HB_effect}
\end{figure}

\textbf{Experimental observations:} Figure \ref{fig:indoor_outdoor_measurements} provides our \emph{indoor and outdoor} measurement results, taken simultaneously for RF  and mmWave.  The output of the MUSIC algorithm is given on the top plots, and the important thing to focus on is the correct AoA in each situation. Note that the AoA is different across different observations plotted. Once that AoA is identified, we compare it with the signal strength (bottom plots) we measured along that direction for the mmWave signal generated at the transmitter location as the RF signal. For the LOS situation, for both indoor and outdoor, there is a strong correlation in the angular composition and the strength of signal coming across all angles in RF and mmWave. This observation is in agreement with \cite{nitsche2015steering}. Indeed, in $94\%$ of all measurements, we have identified the AoA predicted by MUSIC within a $\pm 10\degree$  accuracy for the AoA of the mmWave signal. As a result, based on RF measurements, the correct mmWave transmitter location can be almost perfectly identified under LOS. From Fig. \ref{fig:indoor_outdoor_measurements}, it is evident that as we lose the LOS, the RF/mmWave correlation is lost
and the signal strength in mmWave starts to drop rapidly. However, depending on the size and the location of the
blocker, AoA estimation accuracy varies. For instance, for
a small/mid-size blocker in the middle, in $55\%$ of the observations do the RF and mmWave signals have
their strongest paths within $\pm 10\degree$  of each other. 
In this context, Fig. \ref{HB_effect} demonstrates the effect of human blocker located in the middle compared with when the blocker moves very close to the receiver. From the results, we note that as the blocker moves towards the receiver, the correlation decreases. Moreover, the mmWave signal strength drops when the human blocker is close to the receiver. 

\begin{figure}[t]
\centering
\includegraphics[height=1.5in]{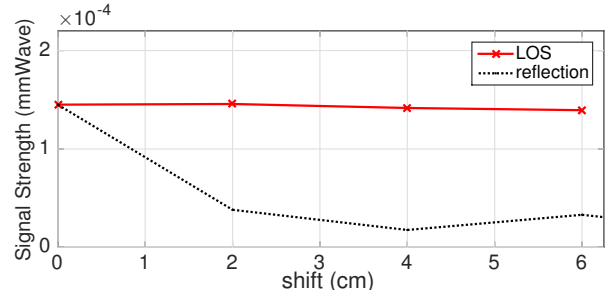}
\vspace{-0.1in}
\caption{\small{Spatial variation in the channel gain. Small movements lead to significant variations without LOS.}}
\label{fig:mobility}
\end{figure}

From the experimental results, our major observation is that in LOS situations, there is a high correlation between the observed RF and mmWave signals, both in signal strength and AoA. Therefore, LOS instances should be exploited in mmWave as much as possible, since there is an associated $10-15$ dB channel gain improvement as well. In order to detect LOS situations,  Fig. \ref{fig:mobility} illustrates the spatial variations of the mmWave channel gain in  LOS and reflection situations. We observe that the LOS situation is quite robust with respect to slight movements, i.e., the large-scale effects lead to minor variations in the channel gain, if the presence of LOS is preserved. On the other hand, if the LOS is blocked and the connection depends on a strong reflector, channel gain becomes relatively unstable and slight movements can result in drastic changes in the channel. As a result,  we use the sensitivity of channel gain to slight movements in order to predict the loss of LOS and take the necessary precautions for a smoother transition in order to mitigate the negative effects of connection losses on the user experience. 
In order to detect LOS situations, the authors in \cite{nitsche2015steering} use the ratio of the highest signal
strength component to the average received signal energy (i.e., peak to average power ratio (PAPR)) as an indicator for LOS inference. As a result, under mobility, the PAPR can be paired with our indicator to boost the LOS detection accuracy. 

\subsection{Beamforming} 
In order to overcome the harsh nature of mmWave channels and compensate for large propagation losses,  highly directional antenna arrays along with beamforming techniques are needed. However, deploying directional antenna arrays makes the cell discovery and access methods costly in terms of delay.  This effect is more pronounced when mobility of users are taken into account where rapid handovers within small-scale cells are needed. Digital beamforming is highly efficient in delay where with the observations from all receive antennas, beamforming can be done by one-shot processing of the observed beacons. However, digital beamforming requires high energy consumption due to the need for a separate ADC per element in the antenna-array. Analog beamforming, on the other hand, is more energy efficient, but it can focus on one direction at a time, making the search process costly in delay. 

\begin{figure}[t]
\centering
\includegraphics[scale=.55]{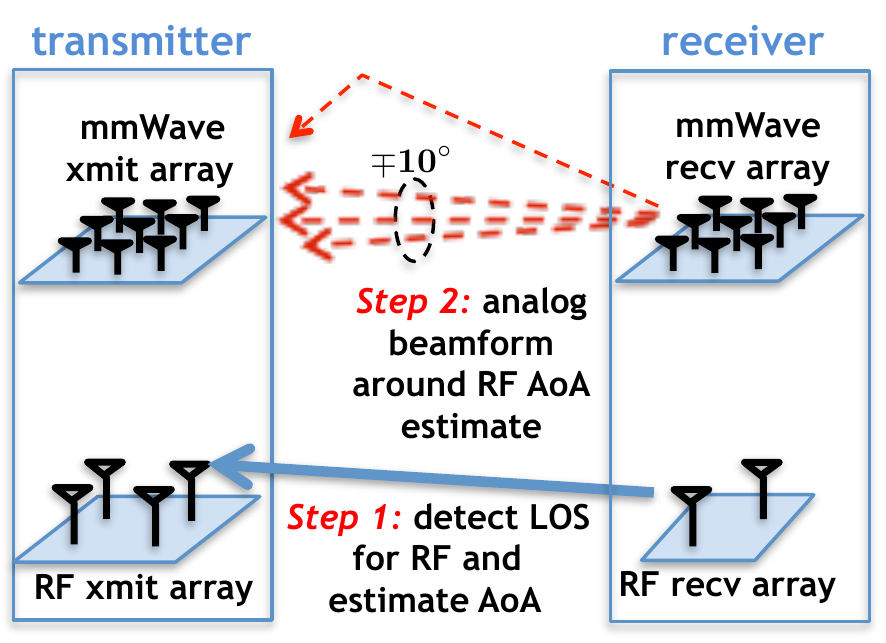}
\caption{\small{Our beamformer works in two phases: (1) the presence of LOS is detected and AoA is estimated all in RF; (2) analog mmWave beamformer focuses on a small area around the estimated AoA.}}
\label{hybrid-beamform}
\end{figure}

Pertaining to the mmWave beamforming efficiency, the mmWave channel is often sparse in the angular domain, with a few scattering clusters, each with several
rays, in addition to a dominant LOS path \cite{andrews2016modeling}. Thus, in order to find the optimal mmWave steering direction $\theta_\text{mm}^*$, our proposed architecture exploits the correlation between the RF and mmWave AoA, and uses a coarse AoA estimation on the RF channel, followed by analog beamforming for fine tuning around the estimated AoA. The RF/mmWave AoA correlation reduces the angular search space and addresses the delay issue of fully-analog beamforming. The RF-assisted mmWave beamforming scheme is specified below, and is graphically illustrated in Fig. \ref{hybrid-beamform}. 

\begin{enumerate}
\item Start the system in the \textbf{RF-only} mode.
\item Implement MUSIC algorithm in RF and estimate the angle of arrival $A_\text{RF}$ based on beacons.
\item Use analog beamforming to fine tune the mmWave beam in the range of $A_\text{RF} \pm 10\degree$:
\begin{enumerate}
\item If the LOS is detected, both interfaces operate jointly in the \textbf{dual RF/mmWave mode} in which resources and arrival traffic are allocated jointly.
\item Otherwise, continue operation of the system in the \textbf{RF-only} mode.
\end{enumerate}
\end{enumerate}

\noindent \textbf{Remark 1:} As our experimental results show, the RF/mmWave correlation decreases as the LOS condition is lost.
However, the RF-assisted beamforming relies on the cross-interface correlation, and once the correlation is lost, it falls back to the traditional beamforming scheme.

\noindent \textbf{Remark 2:} The parameter of searching $\pm 10\degree$ around the estimated AoA is set based on our experimental setup. In general, it will be configured based on dynamics of the scenario and antenna beamwidth. 

 \begin{figure}[t]
\centering
\includegraphics[scale=.65, trim = 1cm 3.1cm 0cm 2.4cm, clip]{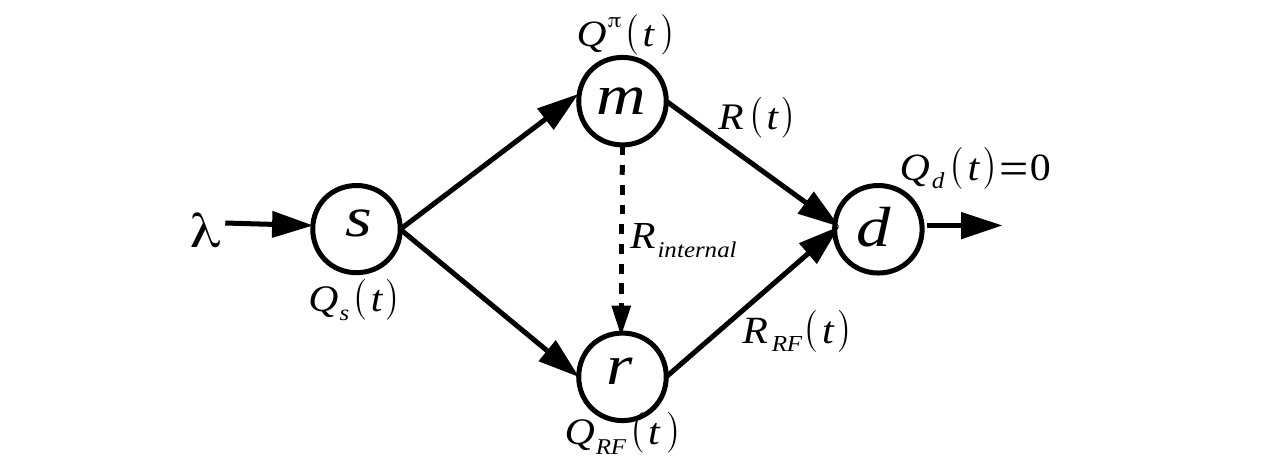}
\caption{An equivalent network model for the integrated RF/mmWave transceiver in which the mmWave (denoted by $m$) and RF (denoted by $r$) interfaces are viewed as individual nodes of the network.}
\label{network-model}
\end{figure}

\section{RF-Assisted mmWave Communications}
\label{rf-assisted-communication}
In the proposed architecture, once the dual RF/mmWave mode is activated, the load division component (in Fig. \ref{fig:system}) schedules the arrival traffic over the RF and mmWave interfaces. The objective is to achieve maximum mmWave throughput with bounded delay performance. To this end, we model our integrated RF/mmWave transceiver as a \emph{diamond network} (see Fig.~\ref{network-model}) in which each of the RF and mmWave interfaces are represented as a network node.  Moreover, a virtual destination (i.e., receiver) node $d$  has been added, and since all data packets are destined for node $d$, its queue length, $Q_d (t)$, is set to be $0$ for all $t$. 
The scheduler node $s$ queries the state information of its neighbor nodes, and assigns packets accordingly. However, due to the high data rate of the mmWave interface, real time tracking of the channel state may not be feasible, and thus, it is desirable to obtain a scheduling policy that is not directly expressed in terms of the CSI. This is in contrast with the classical MaxWeight scheduling policies (e.g., Backpressure) that require CSI information.  

\subsection{Network Model}
\begin{figure}[t]
\centering
\includegraphics[scale=.7, trim = 3.3cm 4cm 3.3cm 3.3cm, clip]{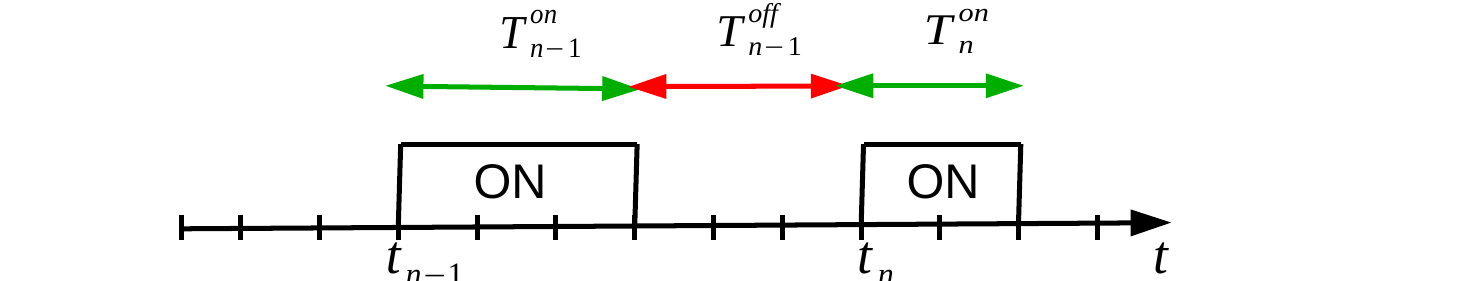}
\vspace{-.2in}
\caption{ON-OFF periods of the mmWave link availability}
\label{on-off-period}
\end{figure}

We assume that the equivalent network model for the integrated RF/mmWave transceiver evolves in discrete (slotted) time $t \in \{0, 1, 2, . . . \}$, and there is an exogenous packet arrival process with rate  $\lambda$. 
To quantify the behavior of the mmWave link using the strongest eigenmode (i.e., corresponding to $\rho_1$), a two-state model (outage and non-outage) can be used. The probability of being in each
state is a function of distance, and statistical models can be fit to characterize it \cite{rappaport2013millimeter}. We use the binary process $\{L(t)\}_{t=1}^\infty$ to account for mmWave outage and non-outage situations such that $L(t) := 1$ implies the availability of the mmWave link (i.e., ON state) during  time slot $t$ and $L(t) := 0$ otherwise (i.e., OFF state). As we also experimentally show in Section \ref{simulation} (see Fig. \ref{fig:channel-mobile}), $L(t) = 1$ corresponds to LOS situations, while $L(t) = 0$ can be mapped to the NLOS situations like human blockers or when there are no strong reflectors. We further assume that $T_n^{\text{on}}$ and $T_n^{\text{off}}$ (with general random variables $T_{\text{on}}$ and $T_{\text{off}}$) denote the $n$-th ON and OFF periods respectively, as shown in Fig. \ref{on-off-period}. 
 The sequence of ON times $\{T_n^{\text{on}}: n \geq 1 \}$ and OFF times $\{T_n^{\text{off}}: n \geq 1 \}$ are independent sequences of i.i.d positive random variables. 
Unlike mmWave, the RF link is much less sensitive to blockage due to diffraction. Thus, for the sake of simplicity, we assume that the RF link is available during all time slots even when $L(t)$ takes on the value of $0$ due to blockers.  

\textbf{State variables and scheduling policy:}
The dynamics of the mmWave link during time slot $t$ is denoted by $\mathbf{x}(t)~=~\big(Q(t), D(t)\big)$ in which $Q(t)$ is the mmWave queue length, and $D(t)$ is the waiting time of the head-of-line packet\footnote{For the sake of notations, we drop the subscript $(.)_\text{mm}$ from the mmWave variables.}. The state space is denoted by $\mathcal{S}$, and  a scheduling policy $\pi \in \Pi$ determines the assignment of packets to the mmWave or RF queue, i.e., $\pi:Q\to\{0, 1\}$ in which $\Pi$ denotes the class of \emph{feasible causal}  policies  in a sense that scheduling decisions are made based on  current state. The decision variable $\pi (Q) = 1$ (or, in short, $\pi = 1$) implies that the packet is routed to the mmWave queue, and $\pi (Q ) = 0$ (or $\pi = 0$) otherwise. The number of packets added to the mmWave queue at time slot $t$ is denoted by $\beta^\pi(t)$.   
To avoid a large waiting time in the mmWave queue due to intermittent connectivity, we require the packets to be \emph{impatient} in the sense that if the waiting time of the head-of-line packet in the mmWave queue exceeds a timeout $T_\text{out}$ (i.e., if $D (t) \geq T_\text{out}$ holds), the packet ``reneges'' (is moved to) to the RF queue. To account for packets reneging, we consider a virtual link between the mmWave and RF queues with a rate equal to the internal read/write speed of processor, as shown in Fig. \ref{network-model}. In this case, $\gamma^\pi(t, T_\text{out})$ denotes the number of reneged packets and $\alpha^\pi(t)$ is the number of packets that are completely served by the mmWave queue. 
Therefore, the  mmWave queue evolves as
$
Q^\pi(t) = \max[0, Q^\pi(t-1) + \beta^\pi(t) - \alpha^\pi(t) - \gamma^\pi(t, T_\text{out})]. 
$

\subsection{Problem Formulation}
\label{problem-formulation}
\begin{definition}{1}\textbf{(Average Throughput and Reneging Rate)} {Given that $\alpha^\pi(t)$ packets are completely served by the mmWave queue and $\gamma^\pi(t, T_\text{out})$ packets renege at time slot $t$, the average throughput and reneging rate of the policy $\pi$ are respectively defined as:}
\begin{equation}
\bar{\alpha}(\pi) = \limsup_{T \rightarrow \infty} \frac{1}{T} \mathds{E} \left[\sum_{t=0}^T \alpha^\pi(t) \right], 
\label{throughput-definition} 
\end{equation} 
\begin{equation}
\bar{\gamma}(\pi, T_\text{out}) = \limsup_{T \rightarrow \infty} \frac{1}{T} \mathds{E} \left[\sum_{t=0}^T \gamma^\pi(t, T_\text{out}) \right].
\label{reneging-rate}  
\end{equation}
\end{definition}

In order for the expectations in \eqref{throughput-definition} and \eqref{reneging-rate} to exist, we assume that $\alpha^\pi(t)$ and $\gamma^\pi(t, T_\text{out})$ are stationary ergodic. 
In this model, imposing the service deadline $T_\text{out}$ ensures that the average waiting time of packets in the mmWave queue is smaller than or equal to $T_\text{out}$. Hence, the reneging mechanism explicitly dictates a constraint on the mmWave waiting time. Therefore, our goal is to derive a throughput-optimal policy with bounded reneging rate. 

\textbf{Problem 1} \textbf{(Constrained Throughput Optimization)} \emph{Given that there is a timeout $T_\text{out}$ for packets in the mmWave queue, we define Problem 1 as follows:
\begin{equation}
\begin{aligned}
 & \max_{\pi \in \Pi}  \ \  \bar{\alpha}(\pi) \\
 & \ \text{s.t.} \ \ \bar{\gamma}(\pi, T_\text{out}) \leq \epsilon \ \ \text{and} \ \ \bar{\beta} (\pi) \leq \lambda,
\end{aligned}
\label{problem-1}
\end{equation}
for a given $\epsilon < \lambda$. }
 
 In \eqref{problem-1}, the objective function and the first constraint can be relaxed as: $\max_{\pi \in \Pi}  \ \  \bar{\alpha}(\pi) - b \big(\bar{\gamma}(\pi, T_\text{out}) - \epsilon\big),$ 
where $b$ is a positive Lagrange multiplier. For any particular fixed value of $b$, it is straightforward to show that there is no loss of optimality in the relaxed problem. To see this, let $\pi^*$ be the optimal policy for the original problem, and $\pi_R^*$ be the optimal policy for the relaxed problem. We have:
\begin{align*}
\bar{\alpha}(\pi^*) \leq \bar{\alpha}(\pi^*) - b \big(\bar{\gamma}(\pi^*, T_\text{out})  - \epsilon\big) \nonumber  \leq  \bar{\alpha}(\pi_R^*) - b \big(\bar{\gamma}(\pi_R^*, T_\text{out})  - \epsilon\big).
\end{align*}
The first inequality holds since $\pi^*$ is feasible in the original problem, and the second inequality holds because $\pi_R^*$ is the optimal solution for the relaxed problem. Thus, there is no loss of optimality in the relaxed problem. 
We note that the relaxed formulation can be interpreted as an optimization over obtained \emph{rewards} and paid \emph{costs}. In particular, each packet that receives service from the mmWave link, results in $r$ units of reward (i.e., in terms of mmWave throughput), while a packet reneging incurs a  cost of $c$ (i.e., in terms of wasted waiting time in the mmWave queue). This leads to the following problem. 

\textbf{Problem 2} \textbf{(Total Reward Optimization)}
\emph{We consider the maximization problem over total rewards obtained as a result of serving packets, and costs due to packets reneging, i.e.,:
\begin{equation}
\begin{aligned}
\ \  \ \ & \max_{\pi \in \Pi} \ \limsup_{T \rightarrow \infty} \frac{1}{T} \mathds{E} \left[ \sum_{t=0} ^T  r \alpha ^ {\pi} (t) - c \gamma^\pi(t, T_\text{out})\right] \\
 & \ \text{s.t.}  \ \ \  \limsup_{T \rightarrow \infty} \frac{1}{T} \mathds{E} \left[ \sum_{t=0} ^T \beta^\pi(t) \right] \leq \lambda,
\end{aligned}
\label{payoff-optimization}
\end{equation} 
where the constraint $\bar{\alpha}(\pi) \leq \bar{\beta}(\pi)$ is implicit. }

\begin{figure}[t]
\centering
\vspace{-.5cm}
\includegraphics[scale=.35]{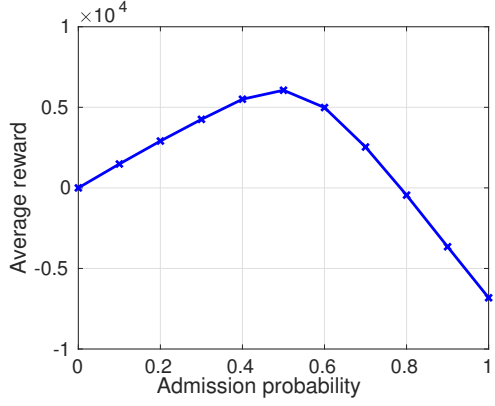}
\vspace{-.3cm}
\caption{Probabilistic admission policy where the objective value of \eqref{payoff-optimization} first increases by admitting more packets into the mmWave queue, and thereafter it decreases due to dominant reneging cost.}
\label{probabilistic}
\end{figure}

It is straightforward to show  that an optimal solution $\pi^*_R$ for the relaxed formulation of Problem 1 is the optimal solution for \eqref{payoff-optimization} and vice versa. To see this, assume that the set of feasible solutions for \eqref{problem-1} is denoted by $\Pi$. For all $\pi \in \Pi$, we have:
$
\bar{\alpha}(\pi) - b \big(\bar{\gamma}(\pi, T_\text{out})\big) \leq \bar{\alpha}(\pi^*_R) - b \big(\bar{\gamma}(\pi^*_R, T_\text{out})\big).
$
Multiplying both sides by $r \geq 0$, and setting $c := rb$, we conclude that $\pi^*_R$ is the optimal solution for  \eqref{payoff-optimization} as well since their feasible sets are identical. 
For $r=1$ and $c=b$, two formulations will be identical. In general, the values of $r$ and $c$ are set based on the application and performance requirements. For instance, a large value of $r$ ensures high throughput, while a large value of  $c$ prioritizes low-latency performance (i.e., a \emph{conservative policy}).
Therefore, \eqref{payoff-optimization} captures the tradeoff between full exploitation of the mmWave capacity and the delay for mmWave channel access through the control knob $\beta^\pi(t)$: if $\beta^\pi(t)$ is set to a very small value for all time slots $t$ (i.e., a conservative policy) then $\alpha^\pi(t)$ would be small as well, and the objective function reduces due to the first term. On the other hand, if $\beta^\pi(t)$ is set to a large value (e.g., matched to the arrival rate $\lambda$ for all time slots $t$) and the link state fluctuates according to the process $\{L(t)\}_{t=1}^\infty$, then the objective function could decrease due to the reneging cost that is captured by the second term. Therefore, there is an optimal value of $\beta^\pi(t)$ within these two extreme cases that results in the maximum return rate. The following example illustrates this point clearly.
 
\textbf{Example (Probabilistic Admission Policy)}: Using a probabilistic admission policy $\pi$, the input arrival rate $\lambda$ is mapped to the admitted rate $\beta^\pi(t)$ such that a packet is admitted to the mmWave queue with an admission probability $p$. We assume that the arrival process is a batch arrival such that there is a batch of size randomly distributed with normal distribution with mean $20$ and standard deviation $1$. The probability of a batch arrival within a time slot is set to $0.9$. The length of ON and OFF periods of the mmWave channel is set according to the log-normal distributions with mean $20$ and $3.5$, respectively.  Fig. \ref{probabilistic} demonstrates behavior of the objective function in \eqref{payoff-optimization} as $p$ increases. We observe that the objective value increases by admitting more packets into the queue up to a certain threshold, and thereafter the objective value decreases due to the dominant reneging cost. Next, we tackle the problem of deriving an optimal admission policy.

\subsection{Optimal RF/mmWave Scheduling Policy }
\label{solution-approach}
From \eqref{payoff-optimization} and using the Lagrangian relaxation, we define: 
\small{
\begin{align}
g(W) = & \max_{\pi \in \Pi} \bigg[ r \bar{\alpha}(\pi) - c \big(\bar{\beta}(\pi) - \bar{\alpha}(\pi)\big) + W \big(\lambda-\bar{\beta}(\pi) \big) \bigg]  
 = & \max_{\pi \in \Pi}  \bigg[(r + c)\bar{\alpha}(\pi) + (W+c) \big(\lambda - \bar{\beta}(\pi)\big) \bigg] - c \lambda,
\label{lagrange}
\end{align}}
\normalsize
\noindent in which the Lagrange multiplier $W$ is positive, and it can be interpreted as a \emph{subsidy} for taking the \emph{passive action}.  In our problem,  passive action is defined as adding packets to the RF queue, while  active action corresponds to admitting packets into the mmWave queue. Hence, the objective is to maximize the long-term expected reward by balancing the reward for serving and the subsidy for passivity.

 The solution of \eqref{lagrange} partitions the state space $\mathcal{S}$ into three sets, $\mathcal{S}_0$, $\mathcal{S}_1$ and $\mathcal{S}_{01}$, where, respectively, the optimal action is $\pi (\mathbf{x}) = 0$ for $\mathbf{x} \in \mathcal{S}_0$, $\pi (\mathbf{x}) = 1$ for $\mathbf{x} \in \mathcal{S}_1$, or some randomization between  $\pi (\mathbf{x}) = 0$ and $\pi (\mathbf{x}) = 1$ for $\mathbf{x} \in \mathcal{S}_{01}$. From \cite{puterman2014markov}, it is known that in a Markov Decision Process if the state space contains a finite number of states, then set $\mathcal{S}_{01}$ does not contain more than one state. This case holds in our model since the mmWave queue length is upper-bounded, and the waiting time in the queue can be at most $T_\text{out}$. Thus, set $\mathcal{S}_{01}$ does not contain more than one state. The following theorem states that Problems 1 and 2 are solved by a \emph{monotone} policy, where a class of policies $\Pi$ has monotone structure if for $\pi \in \Pi$, there exists $h^* \in \{1, 2, ..\}$ such that:
$\pi (Q) = 0 \Longleftrightarrow Q \geq h^*, $ where $Q$ denotes the mmWave queue length. In other words, \emph{the optimal policy routes a packet to the mmWave queue if and only if the mmWave queue length is smaller than an optimal threshold $h^*$.  }

\begin{theorem}(Optimality of monotone policy)
The solution for the reward optimization in \eqref{lagrange} has a monotone structure. 
\label{monotone}
\end{theorem}
\begin{proof}
Let us denote by $v(Q, D)$, the value function corresponding to Problem 2 when mmWave is at state $(Q,D)$, and $V(Q) = \sum_{1 \leq D \leq T_\text{out}} v(Q, D)$. From the Bellman equation \cite{puterman2014markov}, we have:
\begin{align}
 g(W) = V(Q) + \max \bigg\{\lambda V(Q+1) + \sigma Q V(Q-1) + \mu V(Q-1), W + \sigma Q V(Q-1) \nonumber \\ + \mu V(Q-1) \bigg\},
\end{align}

\noindent in which, $\sigma$ is the  reneging rate and $\mu$ is the mmWave service rate. We prove that if passive action is optimal in $Q$ then passive action is optimal in $Q' \geq Q$. Similar to \cite{larranaga2015efficient}, let us define:
\begin{align}
f(Q, 0) & =  r + W + \mu V(Q-1) + \sigma Q V(Q-1); \nonumber\\
f(Q, 1) & =  r +  \lambda V(Q+1) + \sigma Q V(Q-1) +  \mu V(Q-1) \nonumber, 
\label{active}
\end{align}

\noindent and $\varphi (Q) =  \argmax_{a \in \{0,1\}} f(Q,a)$. It then suffices to show that $\varphi (Q') \leq \varphi (Q)$ for $Q' \geq Q$. Assuming $a \leq \varphi (Q')$, we have
$f(Q', \varphi (Q')) - f(Q', a) \geq 0$. Let us now prove that $V(Q)$ has the \emph{subadditivity} property.  

\begin{definition}{2}\textbf{(Subadditive Function)}
Let $X$ and $Y$ be partially ordered sets and $u(x,y)$ a real-valued function on $X \times Y$. We say that $u$ is subadditive if for $x^+ \geq x^-$ in $X$ and $y^+ \geq y^-$ in $Y$ we have: 
$
u(x^+, y^+) + u(x^-, y^-) \leq u(x^+, y^-) + u(x^-, y^+).
$
\end{definition}
To prove that $V(Q)$ is a subadditive function, it suffices to show that for all $Q' \geq Q$ and $a\in \{0,1\}$, the inequality
$f(Q',a) + f(Q, \varphi (Q')) \leq f(Q', \varphi (Q')) + f(Q,a)$ holds. 
If $\varphi (Q') = a = 0$ or $\varphi (Q') = a =1$, then the inequality is satisfied. If $\varphi (Q') = 1$ and $a=0$, then we  show that 
$f(Q,1) - f(Q,0) \leq f(Q',1) - f(Q',0)$, or equivalently, $\lambda V(Q+1) \leq \lambda V(Q'+1)$. This inequality is true due to the fact  that $V(.)$ is non-decreasing, and thus the theorem statement follows. Note that $V(Q)$ is a function of reward and channel state, and it is proportional to the number of packets in the mmWave queue.
\end{proof}
Intuitively, for a first-in-first-out (FIFO) queue, the likelihood that an admitted packet reneges before receiving service increases with the number of queued packets. Therefore, given that the reneging and moving packets from the mmWave queue to the RF queue incurs a delay cost, it is in the scheduler interest to exercise admission control and deny entry to packets when the mmWave queue grows and becomes larger than a threshold. Next, we characterize the value of optimal threshold. 

%
%

\subsection{Optimal Threshold} 
Optimal policy $\pi^*$ imposes a threshold $h^* \in \{0, 1, 2, ..\}$ such that $\pi^*(Q)~=~1$ if and only if $Q < h^*$. Under the ergodicity assumption, we rewrite Problem 2 as:
\begin{equation}
\max_{h\in \{0, 1, 2, ..\}} \bigg(\big(r+c\big) \mathds{E}[\alpha_h] - \big(W+c\big)\mathds{E}[\beta_h]\bigg).
\end{equation}

\begin{lemma}
\label{non-decreasing}
Given an admission threshold $h$, if
\begin{equation}
\psi(h) :=  \frac{\mathds{E}[\beta_{h}] - \mathds{E}[\beta_{h-1}]}{\mathds{E}[\alpha_{h}] - \mathds{E}[\alpha_{h-1}]}, 
\label{decreasing-exp} 
\end{equation}
then $\psi(h)$ is non-decreasing in $h$. 
\end{lemma}

\begin{proof}
In order to prove that $\psi (h)$ is non-decreasing, we note that both $\mathds{E}[\alpha_{h}]$ and $\mathds{E}[\beta_{h}]$ are increasing in $h$ since a larger threshold $h$ results in  admitting more packets (i.e., a larger $\mathds{E}[\beta_{h}]$) and potentially a higher throughput $\mathds{E}[\alpha_{h}]$. Moreover, $\mathds{E}[\beta_{h}]$ is assumed to be an affine function of $h$.  In order to prove that $\psi(h)$ is non-decreasing in $h$, we need to show that $\psi(h+1) - \psi(h) \geq 0 $ for $h \geq 0$. Therefore, we have:
\begin{equation}
\psi(h+1) - \psi(h)  = \frac{\left(\mathds{E}[\alpha_{h}] - \mathds{E}[\alpha_{h-1}]\right)\left(\mathds{E}[\beta_{h+1}] - \mathds{E}[\beta_{h}]\right) - \left(\mathds{E}[\beta_{h}] - \mathds{E}[\beta_{h-1}]\right)\left(\mathds{E}[\alpha_{h+1}] - \mathds{E}[\alpha_{h}]\right)}{\left(\mathds{E}[\alpha_{h+1}] - \mathds{E}[\alpha_{h}]\right)\left(\mathds{E}[\alpha_{h}] - \mathds{E}[\alpha_{h-1}]\right)}.
\label{phi-non-increasing}
\end{equation}
Due to the fact that $\mathds{E}[\alpha_{h}]$ is an increasing and concave, and $\mathds{E}[\beta_{h}]$ is an increasing and affine function of $h$, we conclude that \eqref{phi-non-increasing} is non-negative, and thus $\psi(h+1) - \psi(h) \geq 0$ for $h \geq 0$. 
\end{proof}
%
%
\begin{theorem}
\label{optimal-threshold}
Given an admission threshold $h$, we define 
\begin{equation}
\phi(h) := (W+c) \psi(h). 
\label{opt-threshold} 
\end{equation}
\normalsize
If $\phi(h) < r+c \leq \phi(h+1)$, then $h^* = h$.      
\end{theorem}

\begin{proof}
From Lemma \ref{non-decreasing}, we conclude that $\phi(h)$ is non-decreasing in $h$ as well, i.e., $\phi(h+1) \geq \phi(h), \forall h~\geq~0 $.  For a given threshold $h$ that satisfies $r+c \leq \phi(h+1)$, we conclude that: $$(r+c) \mathds{E}[\alpha_{h+1}] - (W+c) \mathds{E}[\beta_{h+1}] \leq (r+c) \mathds{E}[\alpha_{h}] - (W+c) \mathds{E}[\beta_{h}].$$ Therefore, $h$ achieves a higher objective value than $h+1$. Now in order to establish this result for $h+2$, we can show that: 
$$
r+c \leq \phi(h+1) \leq \phi(h+2) \leq (W+c) \frac{\mathds{E}[\beta_{h+2}] - \mathds{E}[\beta_h]}{\mathds{E}[\alpha_{h+2}] - \mathds{E}[\alpha_{h}]}, 
$$ 
from which we conclude that $h$ is optimal with respect to $h+2$ as well. By induction, we extend this result for all $h' > h$. Similarly, based on the constraint $\phi(h) < r+c$ we prove that $h$ is optimal with respect to all $h' < h$ as well. Thus, $h$ is optimal in general,  and  we have $h^*=h$.  Note that \small{$\mathds{E}[\beta_h] = \lambda \sum_{Q < h, D} \xi_{(Q, D)}$} and \small{$\mathds{E}[\alpha_h] = \mathds{E}[\beta_h] - \sigma \sum_{Q, D = T_\text{out}} \xi_{(Q, D)}$}\normalsize, where $\xi_{(Q, D)}$ denotes the limiting probability of the state $\mathbf{x} = (Q, D)$. Calculation of the limiting distribution is presented in Appendix A. 
\end{proof}

\begin{figure}[t]
\centering
\includegraphics[scale=.25, trim = 0cm 1.25cm 0cm 1.5cm, clip]{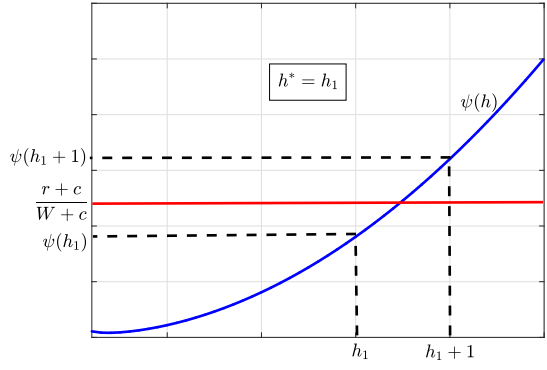}
\caption{A sample path of $\psi(h)$ function and finding the optimal admission threshold.}
\label{threshold_behavior}
\end{figure}

\begin{theorem}
The optimal threshold $h^*$ is an increasing function of $r$ and a decreasing function of $c$. 
\label{thm:optimality-behavior}
\end{theorem}

\begin{proof}
We note that $\frac{r+c}{W+c}$ is increasing in $r$ and decreasing in $c$ due to the fact that $W \leq r$. The condition $W \leq r$ is necessary in order to avoid the trivial scenario where the subsidy is larger than the reward of successful transmission. The trivial scenario leads to always choosing the passive action, and thus we pose the constraint $W \leq r$ to avoid the trivial condition.  From Lemma \eqref{non-decreasing}, we note that $\phi(h)$ is non-decreasing in $h$. From Theorem \ref{optimal-threshold} and because  $\frac{r+c}{W+c}$ is an increasing function of $r$ and a decreasing function of $c$, we conclude that the optimal threshold $h^*$ increases in $r$ and decreases in $c$ as well.  
\end{proof}

The above theorem shows that if the value of $r$ increases, throughput performance will have a higher priority than delay, and thus optimal threshold increases, as expected. On the other hand, by increasing the value of $c$, the optimal threshold decreases to avoid high reneging costs.  As a result, based on the performance requirements, the tradeoff between full exploitation of the mmWave capacity and the delay for mmWave channel access is adjusted through the use of parameters $r$ and $c$.

\subsection{Online Scheduling Policy}
In the previous section, we derived the optimal scheduling policy along with the optimal admission threshold. In practice, the mmWave link is highly dynamic such that the data rate can vary over two orders of magnitude, and thus it is desirable to 
be able to adjust the admission threshold on-the-fly and accommodate the dynamics  of the mmWave channel. In what follows, we provide an online scheduling policy that preserve the form of optimal policy, while  adjusts the admission threshold on-the-fly. 
In order to obtain the online algorithm of Theorem \ref{optimal-threshold}, we note that the optimal threshold $h^*$ is a function of the ratio $ \frac{r+c}{W+c}$ with $W\leq r$. Moreover, the optimal threshold is expressed in terms of function $\psi (h)$ that is an increasing function of $h$. As an example, Fig. \ref{threshold_behavior} demonstrates a sample path of the $\psi(h)$ function introduced in Lemma 1.  In order to calculate the optimal threshold $h^*$ at time $t$, we consider the value of function $\psi(h)$ up to time $t$ and adjust the optimal threshold $h^*$ accordingly (as shown in Fig. \ref{threshold_behavior}).  Algorithm 1 provides an online scheme to calculate the optimal threshold.

\small
\begin{algorithm}[t]
\caption{Online Threshold-based Scheduling Policy}
\begin{varwidth}{\dimexpr\linewidth-2\fboxsep-2\fboxrule\relax}
\begin{algorithmic}[1]
\State $t\gets 1$   \Comment{Set the time to $1$}
\State $h^*(t) \gets K$ 
\Comment{Set $h^*$ equal to mmWave buffer size}
\State $Q(t) \gets 0$      \hspace{1cm}   \Comment{$Q(t):$ mmWave queue length at time $0$}
\State $Q_\text{RF}(t) \gets 0$   
\Comment{$Q_\text{RF}(t):$ RF queue length at time $0$}
\While{$Q_s(t) \neq 0$}     \Comment{Continue until there is no packet}
\If {$Q(t) \leq h^*(t)$}
\State Set $\pi = 1$   \Comment{Add the packet to the mmWave queue}
\Else
\State Set $\pi = 0$   \Comment{Add the packet to the RF queue}
\EndIf
\State Update $Q_s(t), Q_\text{RF}(t), Q(t)$, $\bar{\alpha}(t)$ and $\bar{\beta}(t)$
\State $h^*(t+1)$ = \textsc{Update-Threshold} $\big(\bar{\alpha}(t), \bar{\beta}(t), \bar{\alpha}(t-1), \bar{\beta}(t-1) ,h^*(t)\big)$
\EndWhile

\State
\Function{Update-Threshold} {$\alpha (t),  \beta(t),  \alpha (t-1)$, $\beta (t-1)$, $h (t)$}
\State Calculate $\psi (t) = \frac{\beta (t) - \beta (t-1)}{\alpha (t) - \alpha (t-1)}$
\If {$\psi(t) \geq \frac{r+c}{W+c}$}
\State $h (t+1) \gets h (t) - 1 $
\EndIf
\State \textbf{return} $h (t+1)$
\EndFunction

\end{algorithmic}
\end{varwidth}%
\end{algorithm}
\normalsize

\section{Numerical Results}
\label{simulation}
\begin{figure}[t]
\centering
\includegraphics[scale=.3, trim = 0cm .5cm 0cm 1.25cm, clip]{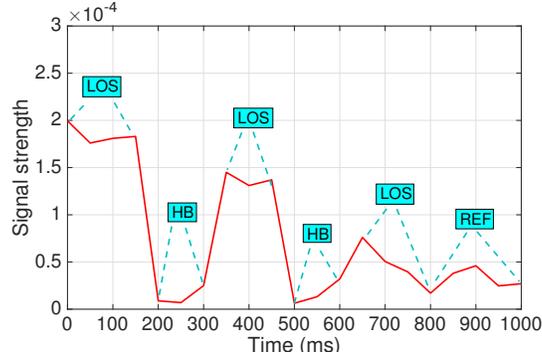}
\vspace{-.3cm}
\caption{\small{MmWave channel temporal correlation under line of sight (LOS), human block (HB), and reflection (REF)}}
\label{fig:channel-mobile}
\end{figure}

In this section, we investigate the performance of our proposed scheduling policy. To this end,  we use the experimental traces to model the ON-OFF mmWave link. In our experiment, a mobile receiver moves with the speed of  $1$ m/s over a path characterized by sudden link transitions due to human blockers (HB) and reflectors (REF). Figure \ref{fig:channel-mobile} illustrates the received signal strength as the mobile moves away from the transmitter. We assume a signal reception cutoff threshold $\delta$ (determined based on the hardware used and environment) such that if the signal strength is below $\delta$, the channel is in the OFF state. Moreover, in order to adequately capture the dynamics of the mmWave channels, the timeout value $T_\text{out}$ is set on-the-fly such that at time $t$, we set $T_\text{out} (t) = \bar{Z}_\text{RF}(t)$ with $\bar{Z}_\text{RF}(t)$ to be the RF average waiting time. Thus, on average, packets would not get stuck in the mmWave queue longer than if they would have joined the RF queue.

\begin{figure}[t!]
  \begin{center}
    \subfigure[Tradeoff between delay and link wastage ratio] {
      \label{tradeoff}
      \includegraphics[scale=.28]{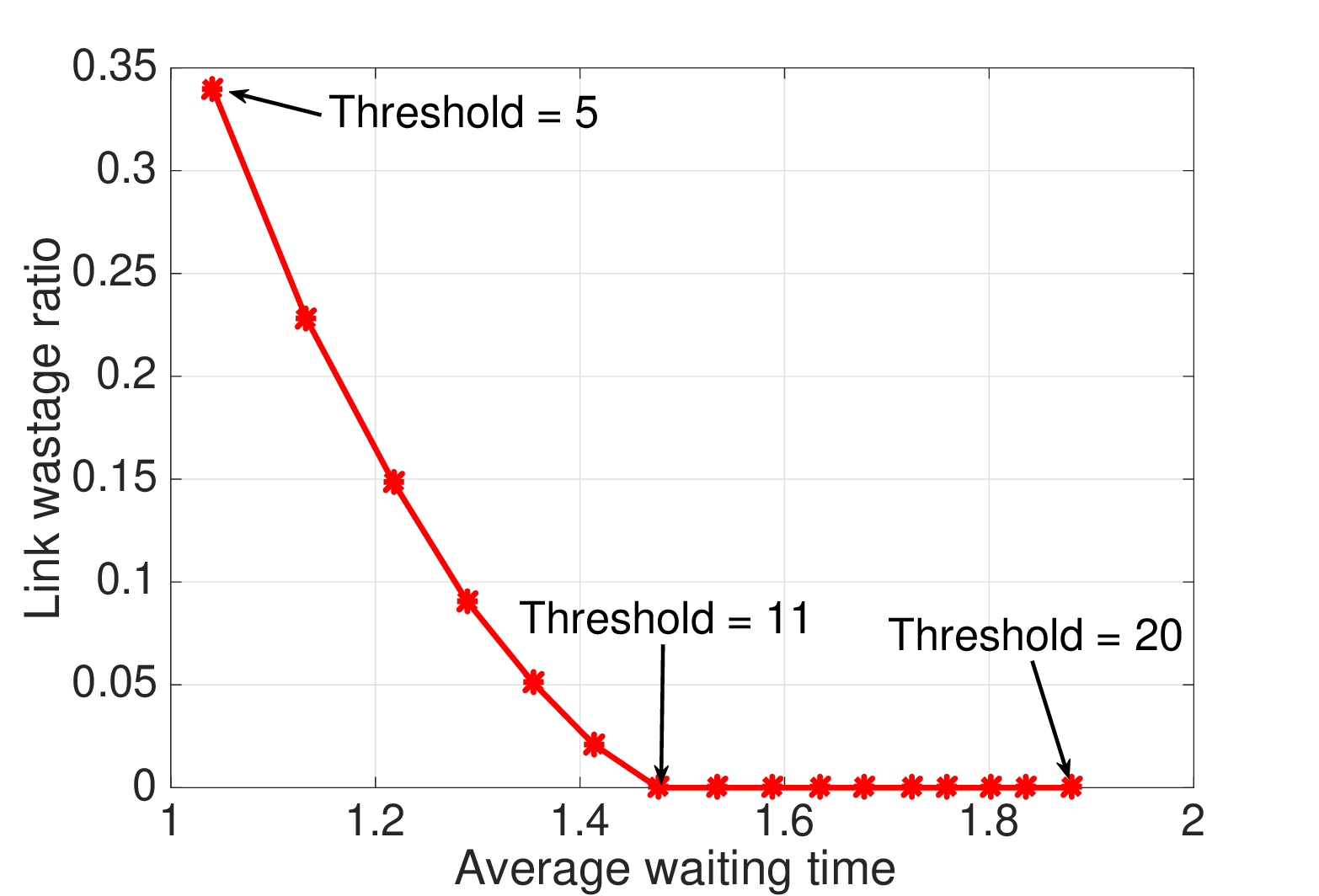} 
    } 
    \subfigure[Total reward obtained as a function of threshold] 
    {
      \label{threshold-based}
      \includegraphics[scale=.28]{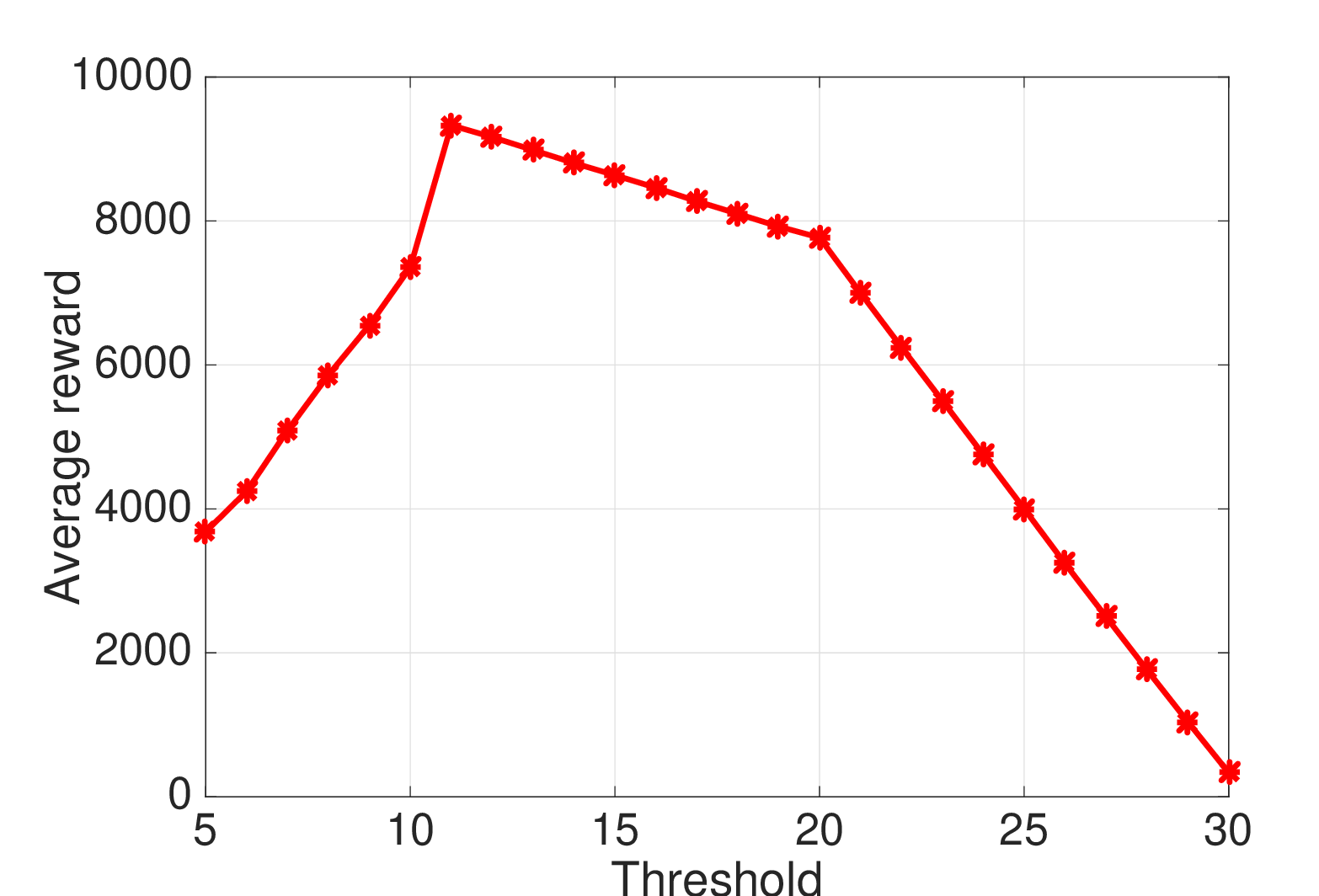}  
    }
  \end{center}
  \vspace{-0.15in}
  \caption{\small{\subref{tradeoff} Trade-off between the average waiting time and link wastage. The control knob is the admission threshold  \subref{threshold-based} Performance of our proposed framework that maximizes total reward. The optimal threshold results in zero wastage and the lowest delay in Fig. \subref{tradeoff}}.}
  \vspace{-0.2in}
  \label{threshold-effect}
\end{figure} 

\subsection{Optimality Results} We first investigate the tradeoff between the mmWave throughput (or, conversely, \emph{link wastage}) and the average waiting time. Link wastage is defined as the fraction of time slots that there are packets in the system, but the mmWave queue is empty and the mmWave link is available (i.e., $L(t) =1$).  
 The tradeoff between link wastage  and the average waiting time is shown in Fig. \ref{tradeoff}. From the results, we observe that if there are so many packets added to the mmWave queue and if the mmWave link becomes unavailable, high average delay incurs. On the other hand, a conservative policy is not desirable either such that due to lack of packets in the mmWave queue, the link wastage increases. Figure \ref{threshold-based} illustrates the total reward obtained as a function of the admission threshold where the maximum reward is obtained for threshold $11$, which is the same threshold value with zero link wastage and the smallest average waiting in Fig. \ref{tradeoff}. 
 
 

%

\begin{figure}[t!]
  \begin{center}
    \subfigure[Throughput performance] {
      \label{throughput-delayedCSI}
      \includegraphics[scale=.25,  trim = .5cm .2cm 0cm .4cm, clip]{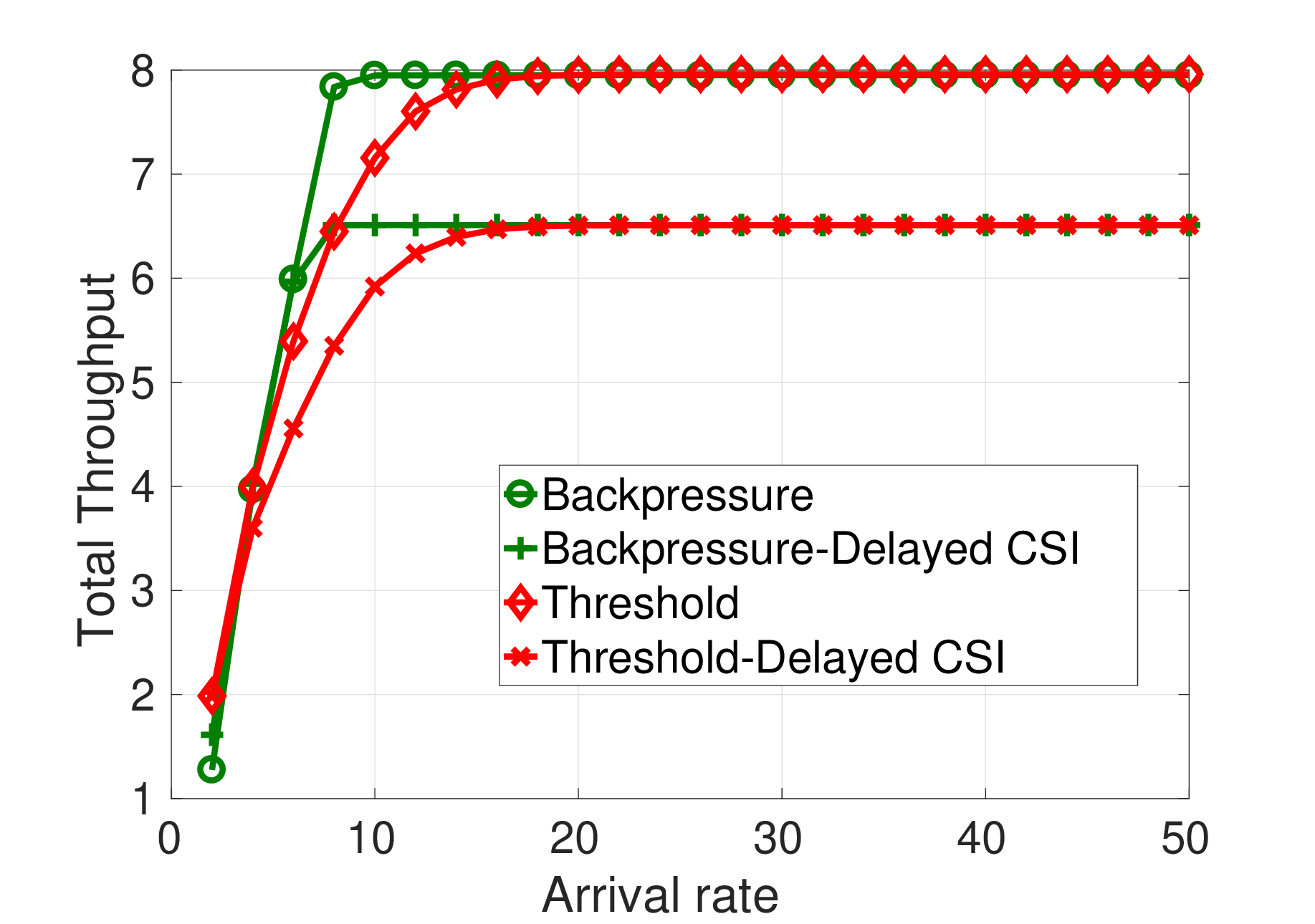} 
    } 
    \subfigure[Delay performance] 
    {\label{delay-delayedCSI}
     \vspace{-.5cm} \includegraphics[scale=.25, trim = .5cm .2cm 0cm .4cm, clip]{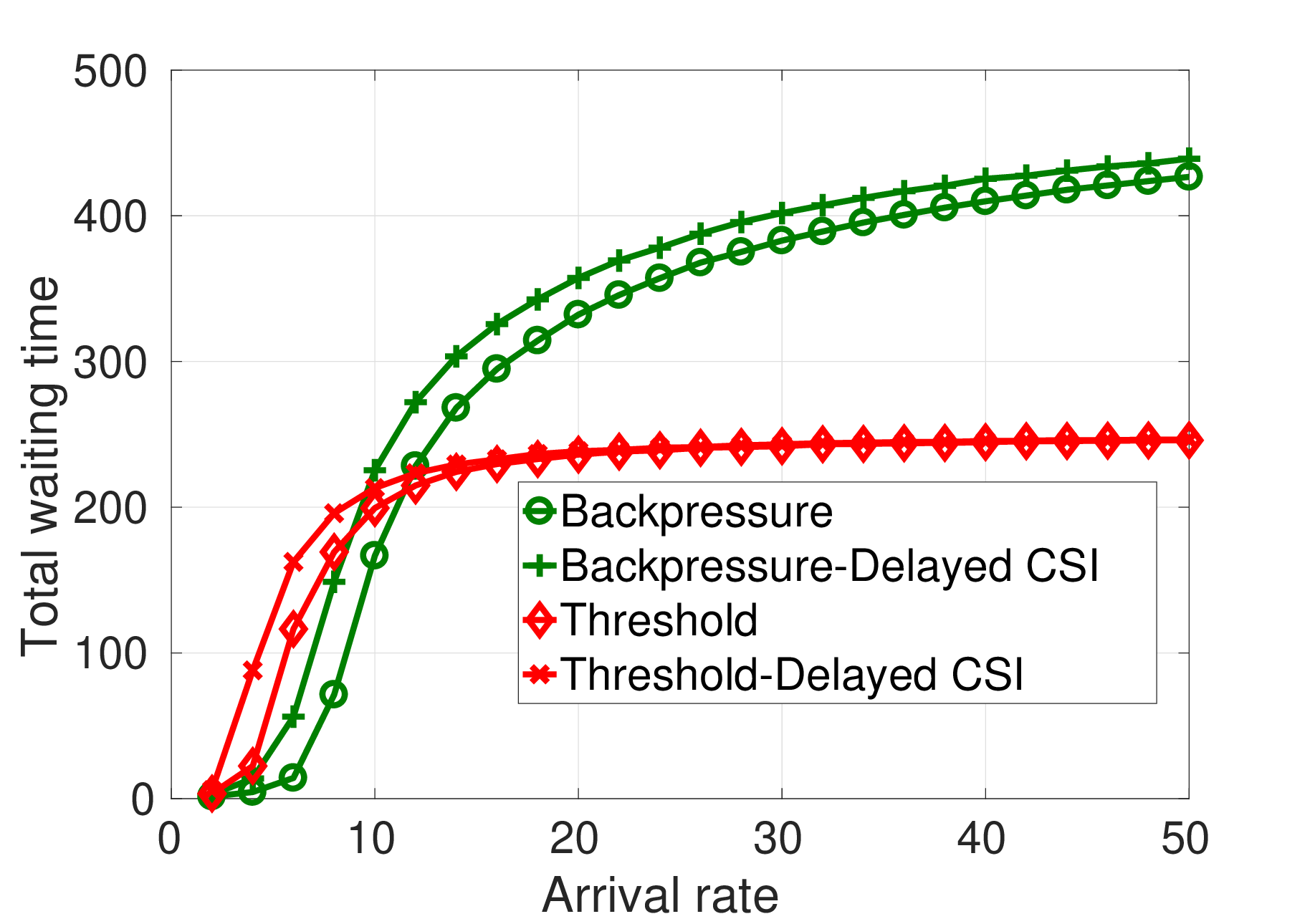}  
    }
  \end{center}
  \vspace{-0.15in}
  \caption{\small{Throughput and delay performance of our proposed threshold-based policy compared with the Backpressure policy under delayed CSI conditions.}}
  \vspace{-0.2in}
  \label{threshold_backpressure_delayedCSI}
\end{figure}

\subsection{Comparison with Backpressure}
In order to optimally design the RF/mmWave transceiver, we represented the transceiver node as a communication network where the RF and mmWave interfaces are modeled as individual network nodes. The objective is to fully exploit the abundant mmWave capacity, while the waiting time is guaranteed to be bounded. In the context of through-optimal scheduling, it is well known that traditional network utility optimization, such as Backpressure policy, promises optimal throughput performance for a wide range of networking problems \cite{georgiadis2006resource}. However, Backpressure policy does not provide any guarantee on the delay performance. Moreover, Backpressure policy requires knowledge of channel state (i.e., link rate), while due to the high data rate of the mmWave interface, real-time tracking of the link state may not be feasible. Therefore, the scheduler node $s$ (in Fig. \ref{network-model}) may obtain information of the data rate of interface $a \in \{ \text{mm}, \text{RF}\}$ with a delay of $\tau_a$. Under the assumption of delayed network state information,  the authors in \cite{ying2011throughput} have shown that the following link selection policy achieves optimal throughput, i.e.,: 
\begin{equation}
a^*(t) = \argmax_{a\in \{\text{mm, RF}\}} \ \big[Q_s(t) - Q_a (t) \big] \mathds{E} [ R_a(t) | R_a(t-\tau_a)], 
\label{delayed-information}
\end{equation}
in which $a^*(t)$ is the optimal interface selected at time $t$, and $R_a(t)$ is the link rate of interface $a$ at time $t$. Under the assumption of delayed CSI, the network stability region (and thus maximum achievable throughput) shrinks as the CSI delay increases \cite{ying2011throughput}.

Figure \ref{throughput-delayedCSI} and \ref{delay-delayedCSI} demonstrate the throughput and delay performance of our threshold-based scheduler compared with the Backpressure algorithm applied for the network model in Fig. \ref{network-model}. We investigate the performance under both real-time CSI and delayed CSI scenarios. 
From the results, we observe that the threshold-based scheme achieves a similar throughput performance to Backpressure. However, delay of Backpressure increases with the arrival traffic, while threshold-based policy provides a bounded delay performance. We note that delayed CSI degrades the performance of both policies, however, the threshold-based policy is more robust (in terms of delay performance) towards CSI delay since it is not \emph{directly} expressed in terms of CSI for scheduling, while the Backpressure policy requires CSI for scheduling.

\subsection{Throughput and Delay Tradeoff}
We showed that the optimal threshold increases by the reward value $r$ and decreases by the reneging cost $c$. As a result, depending on the application requirements (throughput vs. latency), the value of $r$ and $c$ are set, and the optimal threshold value is regulated accordingly. 
\begin{figure}[t!]
  \begin{center}
    \subfigure[Throughput performance] 
    {
      \label{throughput-cost}
      \includegraphics[scale=.25,  trim = .5cm .2cm 0cm .4cm, clip]{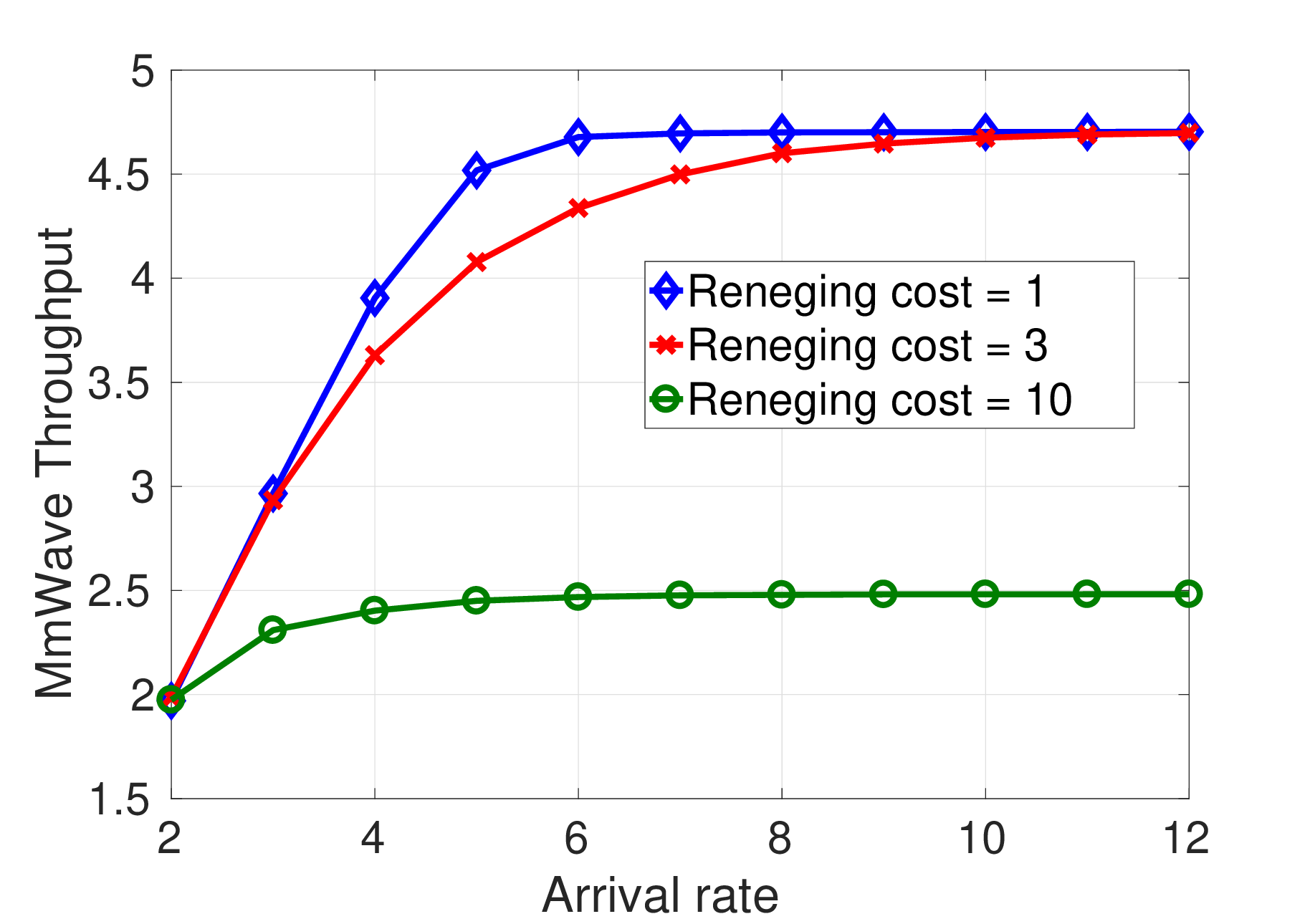} 
    } 
    \subfigure[Delay performance] 
    {\label{delay-cost}
     \vspace{-.5cm} \includegraphics[scale=.25, trim = .5cm .2cm 0cm .4cm, clip]{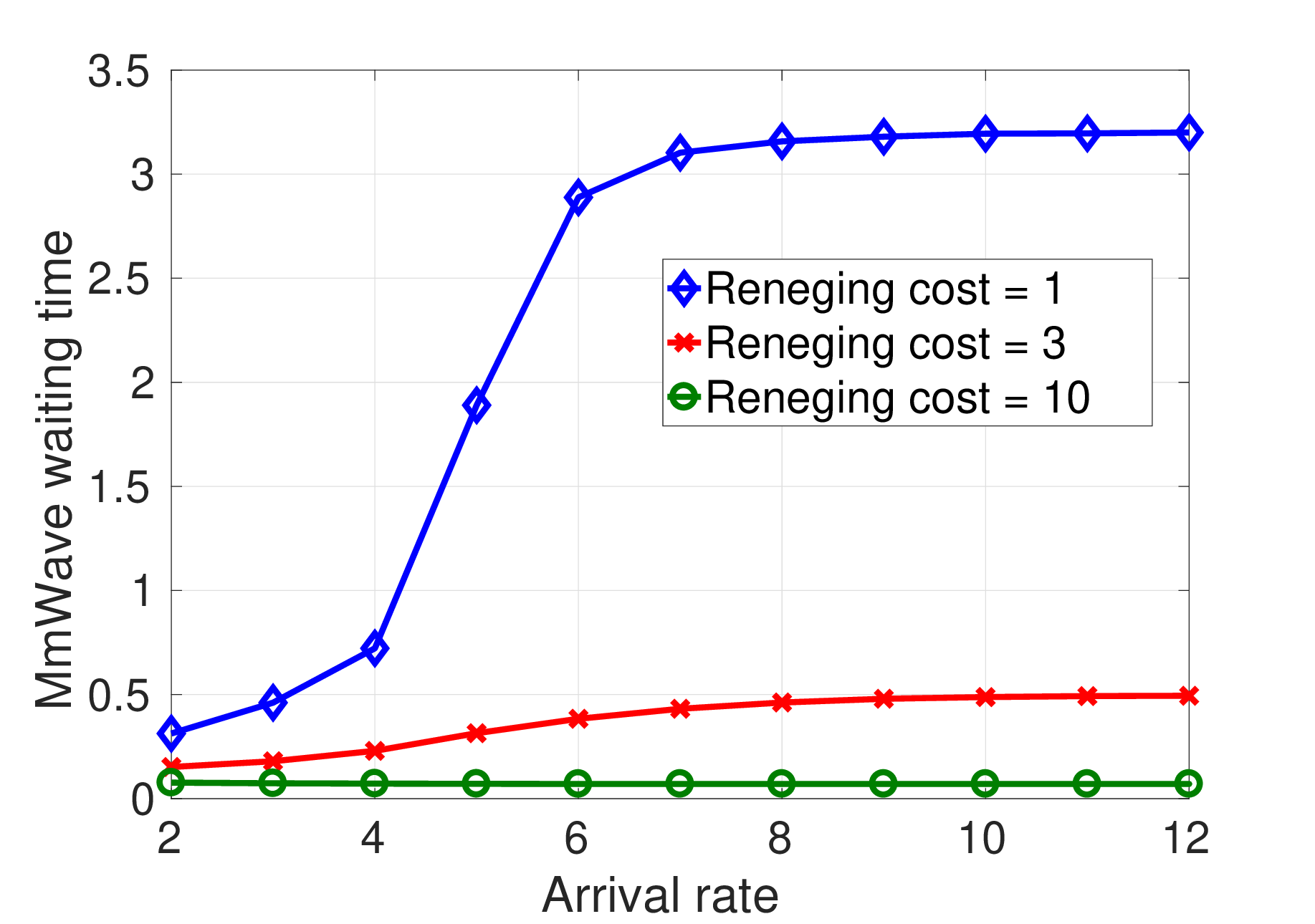}  
    }
    \subfigure[Throughput performance] {
      \label{throughput-reward}
      \includegraphics[scale=.25,  trim = .5cm .2cm 0cm .4cm, clip]{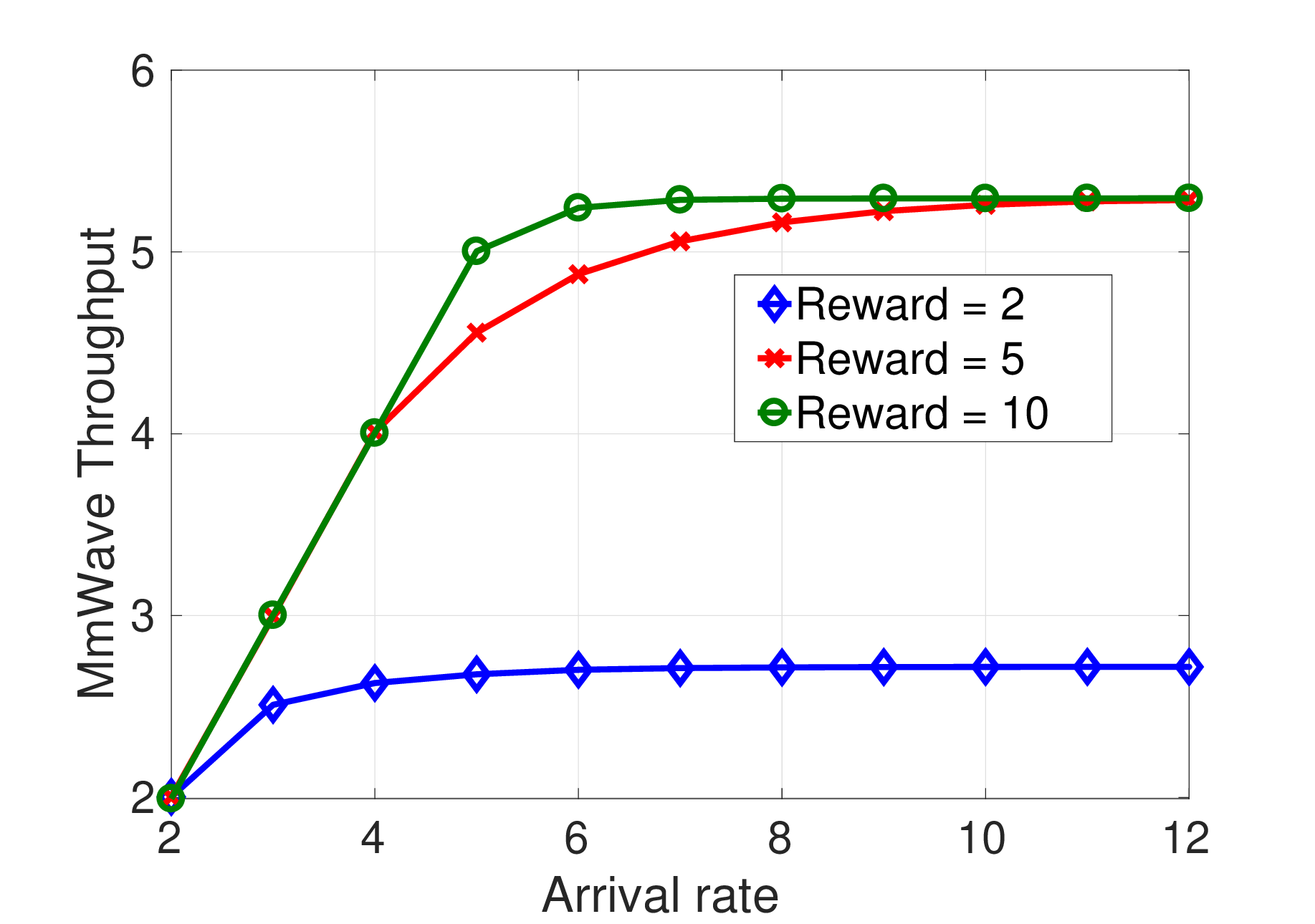} 
    } 
    \subfigure[Delay performance] 
    {\label{delay-reward}
     \vspace{-.5cm} \includegraphics[scale=.25, trim = .5cm .2cm 0cm .4cm, clip]{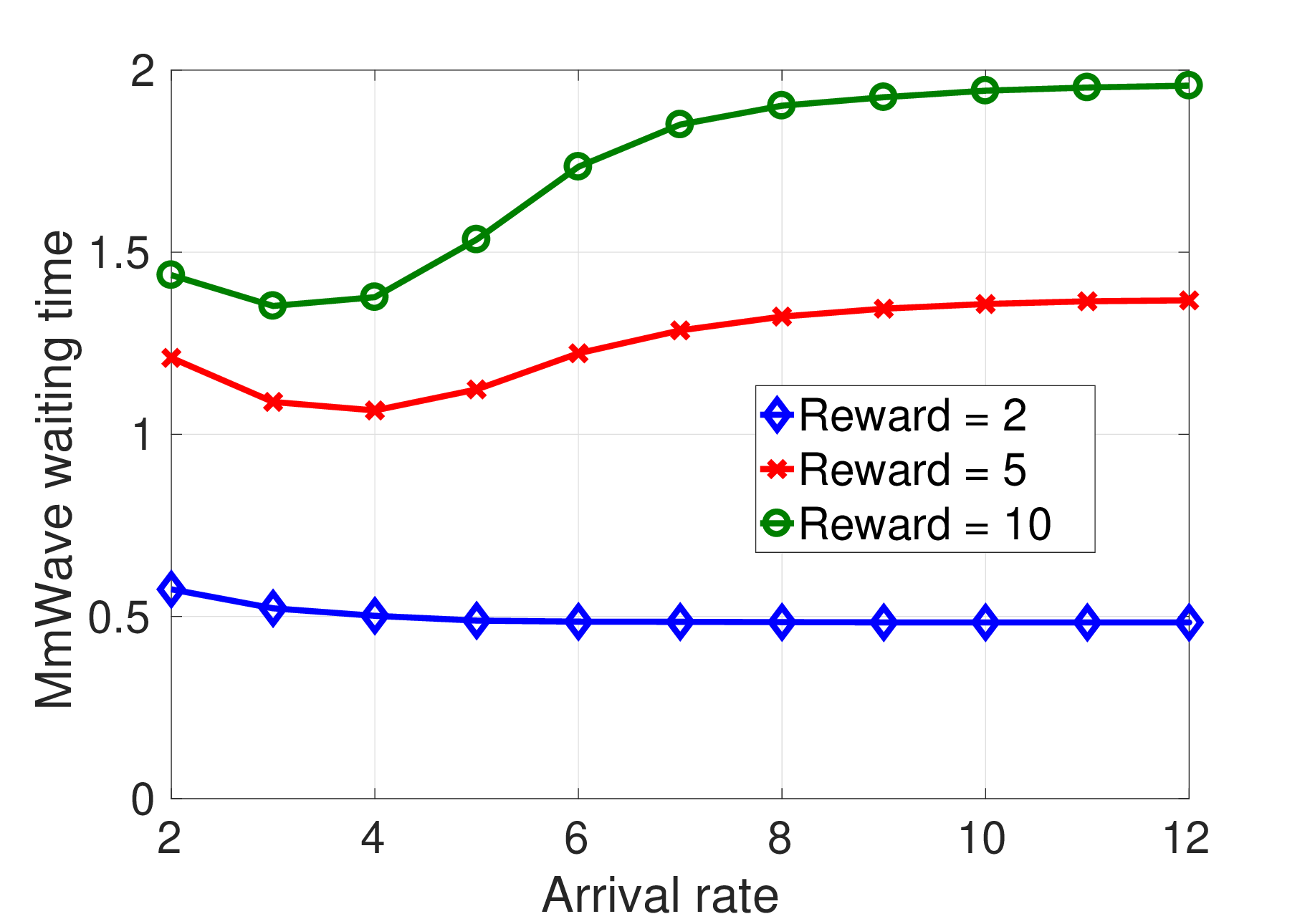}  
    }
  \end{center}
  \vspace{-0.15in}
  \caption{\small{Throughput and delay performance of our proposed threshold-based policy compared with different values of reneging cost.}}
  \vspace{-0.2in}
  \label{threshold_reneging_cost}
\end{figure}
Figure \ref{throughput-cost} and \ref{delay-cost} demonstrate the throughput and delay performance of the threshold-based policy as the reneging cost $c$ increases. From the results, we note that by increasing the value of $c$, the optimal threshold decreases and thus less packets are admitted to the mmWave queue, as expected. As a result, mmWave waiting time decreases while due to the lack of backlogged packets, throughput performance degrades as well. On the other hand, the trade off between the mmWave throughput and waiting time can be balanced by adjusting the value of reward $r$. Figure \ref{throughput-reward} and \ref{delay-reward} illustrate the throughput and delay performance as the reward value $r$ increases. Similarly, as the reward value $r$ increases, the optimal threshold increases and more packets are admitted into the mmWave queue, which results in a higher throughput at the cost of larger waiting time.

\section{Conclusion}
\label{conclusion}
In this paper, we proposed an integrated RF/mmWave architecture for $5$G cellular systems. Our proposed architecture includes an RF-assisted beamforming that exploits the correlation between the RF and mmWave interfaces in order to enhance the energy efficiency of mmWave beamforming. In addition to beamforming,  we utilized the RF interface for data transfer, and formulated an optimal scheduling policy in order to maximize the long-term throughput of the mmWave interface provided that the average delay is bounded. We cast the constrained throughput maximization as a reward optimization, and proved that the optimal scheduling policy has a simple monotone structure. As a result,  using the RF interface as a secondary data transfer mechanism, the abundant yet intermittent mmWave bandwidth is fully utilized. Indeed, we believe that mmWave will most likely be deployed with an overlay of RF in $5$G.

\begin{appendices}
\section{Calculating the Limiting Distribution}
In order to characterize the value of optimal threshold, we calculate the limiting distribution of the state of mmWave queue. 
To this end, the authors in \cite{kim2014analytical} introduced an \emph{embedding technique} such that an embedded process $\{\mathbf{x}_n\}_{n=1}^\infty$ is obtained by sampling the process $\{\mathbf{x}(t)\}_{t=1}^\infty$ at the beginning of each ON period (see \cite{kim2014analytical} for details). We assume that the limiting distribution of the mmWave queue at state $\boldsymbol{i}$ is denoted by $\xi_{\text{off}}^{({\boldsymbol{i}})}$ and $\xi_{\text{on}}^{({\boldsymbol{i}})}$ for $L(t) =0$ and $L(t) =1$, respectively:
\begin{align}
\xi_\text{off}^{({\boldsymbol{i}})} := \lim_{t\rightarrow \infty} (\mathbf{x}(t) = \boldsymbol{i} , L(t) =0); \quad
\xi_\text{on}^{({\boldsymbol{i}})} := \lim_{t\rightarrow \infty} (\mathbf{x}(t) = \boldsymbol{i} , L(t) =1). 
\end{align}
As in \cite{kim2014analytical}, the limiting distribution of all states $\boldsymbol{i} \in {\mathcal{S}}$ under the OFF and ON link state is then obtained in a matrix form as follows:
\begin{align}
\ \ \ \ \ \boldsymbol{\xi}_{\text{off}} & =  \frac{\boldsymbol{\nu} \mathds{E}\big[(\boldsymbol{M}_{\text{on}})^{T_{\text{on}}}\sum_{k = 1}^{T_{\text{off}}}  (\boldsymbol{M}_{\text{off}})^{k-1}\big]}{\mathds{E}\big[T_{\text{on}} + T_{\text{off}}\big]}; \nonumber \\
 \boldsymbol{\xi}_{\text{on}} & =  \frac{\boldsymbol{\nu} \mathds{E}\big[\sum_{k = 1}^{T_{\text{on}}}  (\boldsymbol{M}_{\text{on}})^{k-1}\big]}{\mathds{E}\big[T_{\text{on}} + T_{\text{off}}\big]},
\end{align}
 where $\boldsymbol{\nu}$ is the vector of limiting distribution for the embedded process $
\{\mathbf{x}_n\}_{n=1}^\infty$. Moreover, $\boldsymbol{M}_{\text{off}} = \big[P^{({\boldsymbol{i},\boldsymbol{j}})}_{\text{off}}\big]$ and $\boldsymbol{M}_{\text{on}} = \big[P^{(\boldsymbol{i},\boldsymbol{j})}_{\text{on}}\big]$ such that: 
\begin{align}
P^{(\boldsymbol{i},\boldsymbol{j})}_{\text{off}} := P\big(\mathbf{x}(t+1) = \boldsymbol{j} | \mathbf{x}(t) = \boldsymbol{i}, L(t) = 0\big), \nonumber \\
P^{(\boldsymbol{i},\boldsymbol{j})}_{\text{on}} := P\big(\mathbf{x}(t+1) = \boldsymbol{j} | \mathbf{x}(t) = \boldsymbol{i}, L(t) = 1\big).
\end{align}
The proof is similar to \cite{kim2014analytical}.  Therefore, the limiting distribution vector of the state space $\mathcal{S}$ is obtained as: $\boldsymbol{\xi} = \boldsymbol{\xi}_{\text{off}} + \boldsymbol{\xi}_{\text{on}}$. A sufficient condition for existence of the limiting distribution is that the embedded process has finite state space, which holds in our model due to a bounded queue length and waiting time. Our model involves an admission policy that regulates the arrival process, and thus length of the mmWave queue does not exceed an optimal threshold $h^*$. To denote the limiting distribution at the state $\mathbf{x} = (Q, D)$, we use the notation $\xi_{(Q, D)}$. 

\end{appendices}

\bibliographystyle{IEEEtran}
{\footnotesize\bibliography{../References}}

\begin{thebibliography}{10}
\providecommand{\url}[1]{#1}
\csname url@samestyle\endcsname
\providecommand{\newblock}{\relax}
\providecommand{\bibinfo}[2]{#2}
\providecommand{\BIBentrySTDinterwordspacing}{\spaceskip=0pt\relax}
\providecommand{\BIBentryALTinterwordstretchfactor}{4}
\providecommand{\BIBentryALTinterwordspacing}{\spaceskip=\fontdimen2\font plus
\BIBentryALTinterwordstretchfactor\fontdimen3\font minus
  \fontdimen4\font\relax}
\providecommand{\BIBforeignlanguage}[2]{{%
\expandafter\ifx\csname l@#1\endcsname\relax
\typeout{** WARNING: IEEEtran.bst: No hyphenation pattern has been}%
\typeout{** loaded for the language `#1'. Using the pattern for}%
\typeout{** the default language instead.}%
\else
\language=\csname l@#1\endcsname
\fi
#2}}
\providecommand{\BIBdecl}{\relax}
\BIBdecl

\bibitem{khan2011mmwave}
F.~Khan and Z.~Pi, ``mm{W}ave mobile broadband ({MMB}): Unleashing the
  3--300{GH}z spectrum,'' in \emph{34th IEEE Sarnoff Symposium}, 2011.

\bibitem{rappaport2013millimeter}
T.~S. Rappaport, S.~Sun, R.~Mayzus, H.~Zhao, Y.~Azar, K.~Wang, G.~N. Wong,
  J.~K. Schulz, M.~Samimi, and F.~Gutierrez, ``Millimeter wave mobile
  communications for 5{G} cellular: It will work!'' \emph{Access, IEEE},
  vol.~1, pp. 335--349, 2013.

\bibitem{roh2014millimeter}
W.~Roh, J.-Y. Seol, J.~Park, B.~Lee, J.~Lee, Y.~Kim, J.~Cho, K.~Cheun, and
  F.~Aryanfar, ``Millimeter-wave beamforming as an enabling technology for 5{G}
  cellular communications: theoretical feasibility and prototype results,''
  \emph{IEEE Communications Magazine}, vol.~52, no.~2, 2014.

\bibitem{mo2016hybrid}
J.~Mo, A.~Alkhateeb, S.~Abu-Surra, and R.~W. Heath~Jr, ``Hybrid architectures
  with few-bit {ADC} receivers: Achievable rates and energy-rate tradeoffs,''
  \emph{arXiv preprint arXiv:1605.00668}, 2016.

\bibitem{nitsche2015steering}
T.~Nitsche, A.~B. Flores, E.~W. Knightly, and J.~Widmer, ``Steering with eyes
  closed: mm-wave beam steering without in-band measurement,'' in
  \emph{Computer Communications (INFOCOM), IEEE Conference on}.\hskip 1em plus
  0.5em minus 0.4em\relax IEEE, 2015, pp. 2416--2424.

\bibitem{aliestimating}
A.~Ali, N.~Prelcic, and R.~Heath, ``Estimating millimeter wave channels using
  out-of-band measurements,'' \emph{Information Theory and Applications
  Workshop (ITA)}, 2016.

\bibitem{wiopt-2017}
M.~Hashemi, C.~E. Koksal, and N.~B. Shroff, ``Hybrid {RF-mmWave} communications
  to achieve low latency and high energy efficiency in {5G} cellular systems,''
  in \emph{Modeling and Optimization in Mobile, Ad Hoc, and Wireless Networks
  (WiOpt), 15th International Symposium on}.\hskip 1em plus 0.5em minus
  0.4em\relax IEEE, 2017.

\bibitem{collonge2004influence}
S.~Collonge, G.~Zaharia, and G.~E. Zein, ``Influence of the human activity on
  wide-band characteristics of the 60 {GH}z indoor radio channel,''
  \emph{Wireless Communications, IEEE Transactions on}, vol.~3, no.~6, pp.
  2396--2406, 2004.

\bibitem{rangan2014millimeter}
S.~Rangan, T.~S. Rappaport, and E.~Erkip, ``Millimeter-wave cellular wireless
  networks: Potentials and challenges,'' \emph{Proceedings of the IEEE}, vol.
  102, no.~3, pp. 366--385, 2014.

\bibitem{rappaport2014millimeter}
T.~S. Rappaport, R.~W. Heath~Jr, R.~C. Daniels, and J.~N. Murdock,
  \emph{Millimeter wave wireless communications}.\hskip 1em plus 0.5em minus
  0.4em\relax Pearson Education, 2014.

\bibitem{nurmela2015metis}
V.~Nurmela, A.~Karttunen, A.~Roivainen, L.~Raschkowski, T.~Imai,
  J.~Jarvelainen, J.~Medbo, J.~Vihriala, J.~Meinila, K.~Haneda \emph{et~al.},
  ``{METIS} channel models,'' \emph{Seventh FrameworN Programme ICT-317669},
  2015.

\bibitem{adhikary2014joint}
A.~Adhikary, E.~Al~Safadi, M.~K. Samimi, R.~Wang, G.~Caire, T.~S. Rappaport,
  and A.~F. Molisch, ``Joint spatial division and multiplexing for mm-wave
  channels,'' \emph{IEEE Journal on Selected Areas in Communications}, vol.~32,
  no.~6, pp. 1239--1255, 2014.

\bibitem{ali2017millimeter}
A.~Ali, N.~Gonz{\'a}lez-Prelcic, and R.~W. Heath~Jr, ``Millimeter wave
  beam-selection using out-of-band spatial information,'' \emph{arXiv preprint
  arXiv:1702.08574}, 2017.

\bibitem{kim2014analytical}
Y.~Kim, K.~Lee, and N.~B. Shroff, ``An analytical framework to characterize the
  efficiency and delay in a mobile data offloading system,'' in
  \emph{Proceedings of the 15th ACM international symposium on Mobile ad hoc
  networking and computing}.\hskip 1em plus 0.5em minus 0.4em\relax ACM, 2014,
  pp. 267--276.

\bibitem{jindal2005grouping}
S.~Jindal, A.~Jindal, and N.~Gupta, ``Grouping {Wi-MAX, 3G and Wi-Fi} for
  wireless broadband,'' in \emph{1st IEEE and IFIP International Conference in
  Central Asia on Internet}, 2005.

\bibitem{nitsche2014ieee}
T.~Nitsche, C.~Cordeiro, A.~B. Flores, E.~W. Knightly, E.~Perahia, and J.~C.
  Widmer, ``{IEEE} 802.11 ad: directional {60 GHz} communication for
  multi-gigabit-per-second {Wi-Fi} [invited paper],'' \emph{Communications
  Magazine, IEEE}, vol.~52, no.~12, pp. 132--141, 2014.

\bibitem{shokri2015design}
H.~Shokri-Ghadikolaei, C.~Fischione, P.~Popovski, and M.~Zorzi, ``Design
  aspects of short range millimeter wave networks: A mac layer perspective,''
  \emph{arXiv preprint arXiv:1509.07538}, 2015.

\bibitem{nitsche2015boon}
T.~Nitsche, G.~Bielsa, I.~Tejado, A.~Loch, and J.~Widmer, ``Boon and bane of
  {60 GHz} networks: Practical insights into beamforming, interference, and
  frame level operation,'' 2015.

\bibitem{yildirim2009cross}
F.~Yildirim and H.~Liu, ``A cross-layer neighbor-discovery algorithm for
  directional {60-GHz} networks,'' \emph{Vehicular Technology, IEEE
  Transactions on}, vol.~58, no.~8, pp. 4598--4604, 2009.

\bibitem{mezzavilla20155g}
M.~Mezzavilla, S.~Dutta, M.~Zhang, M.~R. Akdeniz, and S.~Rangan, ``{5G} mmwave
  module for the ns-3 network simulator,'' in \emph{Proceedings of the 18th ACM
  International Conference on Modeling, Analysis and Simulation of Wireless and
  Mobile Systems}.\hskip 1em plus 0.5em minus 0.4em\relax ACM, 2015, pp.
  283--290.

\bibitem{andrews2016modeling}
J.~G. Andrews, T.~Bai, M.~Kulkarni, A.~Alkhateeb, A.~Gupta, and R.~W. Heath~Jr,
  ``Modeling and analyzing millimeter wave cellular systems,'' \emph{arXiv
  preprint arXiv:1605.04283}, 2016.

\bibitem{schmidt1986multiple}
R.~Schmidt, ``Multiple emitter location and signal parameter estimation,''
  \emph{IEEE transactions on antennas and propagation}, vol.~34, no.~3, pp.
  276--280, 1986.

\bibitem{puterman2014markov}
M.~L. Puterman, \emph{Markov decision processes: discrete stochastic dynamic
  programming}.\hskip 1em plus 0.5em minus 0.4em\relax John Wiley \& Sons,
  2014.

\bibitem{larranaga2015efficient}
M.~Larra{\~n}aga, O.~J. Boxma, R.~N{\'u}{\~n}ez-Queija, and M.~S. Squillante,
  ``Efficient content delivery in the presence of impatient jobs,'' in
  \emph{Teletraffic Congress (ITC 27), 27th International}.\hskip 1em plus
  0.5em minus 0.4em\relax IEEE, 2015, pp. 73--81.

\bibitem{georgiadis2006resource}
L.~Georgiadis, M.~J. Neely, and L.~Tassiulas, \emph{Resource allocation and
  cross-layer control in wireless networks}.\hskip 1em plus 0.5em minus
  0.4em\relax Now Publishers Inc, 2006.

\bibitem{ying2011throughput}
L.~Ying and S.~Shakkottai, ``On throughput optimality with delayed
  network-state information,'' \emph{IEEE Transactions on Information Theory},
  vol.~57, no.~8, pp. 5116--5132, 2011.

\end{thebibliography}

\end{document}